# Decoupling Strain-Rate Sensitivity and Deformation Length Scale Effects in Neutron-Irradiated Tungsten: A Coupled Nano-Indentation, HR-EBSD and Crystal Plasticity Study

Prashil S. Joshi[a], Bethany Jim[b], Chris Hardie[c], Angus Wilkinson[b], David EJ Armstrong[b], Suchandrima Das[a*]

[a]*Department of Materials Engineering, Indian Institute of Science, CV Raman Road, Bangalore, 560012, India*

[b]*Department of Materials, University of Oxford, Parks road, Oxford OX1 3PJ, UK*

[c]*UK Atomic Energy Authority, Culham Science Centre, Abingdon, Oxfordshire, OX14 3DB, UK*

Corresponding author email: suchandrima@iisc.ac.in

## Graphical Abstract

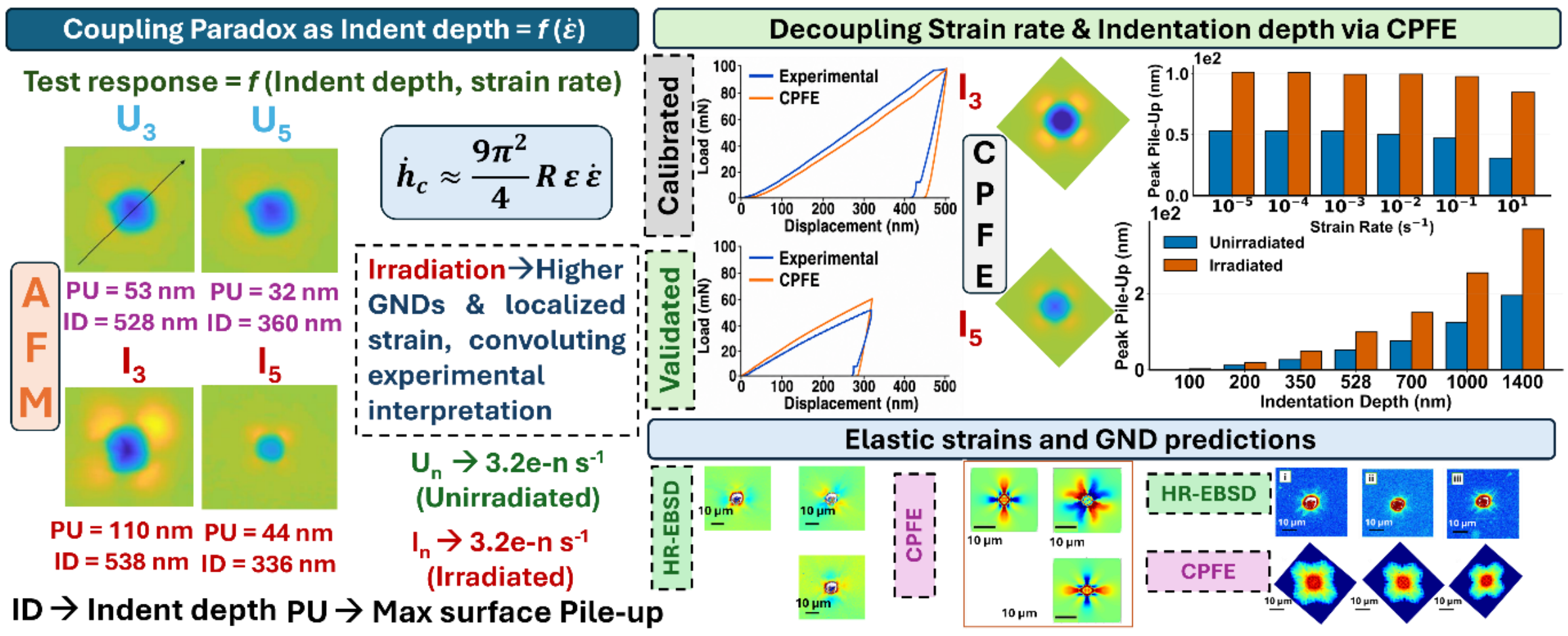

**Abstract**

Strain-rate-dependent deformation in spherical nanoindentation couples material rate sensitivity effects with the evolving contact radius, penetration depth and plastic-zone size. A rate-sensitive model where loading rate and indentation depth are prescribed independently allows their contributions to be disentangled. We study this in unirradiated and neutron-irradiated single-crystal tungsten, indented to the same final indentation strain at nominal strain-rates of $3.2 \times 10^{-5}$ to $3.2 \times 10^{-3}$ $s^{-1}$ and reaching different final depths. Surface pile-up, residual lattice strain and dislocation distributions were measured by AFM and HR-EBSD. A rate-sensitive, strain-gradient crystal plasticity framework, incorporating thermally activated slip and irradiation hardening derived from measured defect populations, was calibrated at the highest strain rate and reproduced all these observables at the two lower strain rates without further parameter adjustment. The framework was then used to vary strain rate and depth independently. At fixed depth, strain rate over the experimental range changes the predicted peak load by < 1% and the pile-up by < 2 nm. At fixed strain rate, the experimental depth differences alone produce pile-up increases of ~ 90% and 130%, comparable to the measured 66% and 150% for unirradiated and irradiated tungsten. The measured differences between strain rate conditions therefore predominantly reflect the different depths reached; a 10% difference in final depth mimics more than an order of magnitude change in strain rate. Irradiation is seen to promote localisation, with its effect on pile-up increasing with depth up to ~ 500 nm, while remaining consistent with a common thermally activated deformation mechanism.

## 1. Introduction

The long-term structural stability of plasma facing component materials (PFC) continues to be a major hurdle for the implementation of fusion energy systems (Loarte et al., 2003; Miranda et al., 2026). While in operation, components that face plasma endure continuous thermal and irradiation stress, in addition to transient occurrences like edge-localised modes (ELMs) and plasma disruptions (Federici et al., 2003; Wirtz et al., 2015). Such events may deposit thermal loads of multiple GW $m^{-2}$ over sub-millisecond intervals, resulting in extreme thermal gradients, localized stresses, and dynamically changing deformation fields (Coenen et al., 2015; Yuan et al., 2016). The resulting mechanical behavior is essential for assessing component reliability, crack formation, and lifespan. Thus, comprehending the PFC's deformation characteristics under different loading rates is crucial for creating physical descriptions of material behavior in fusion settings.

Tungsten is a leading candidate material for plasma-facing components in future fusion reactors due to its remarkably high melting point, advantageous thermal conductivity, minimal sputtering yield, and low retention of tritium (Das et al., 2019a; Pitts et al., 2013; Rieth et al., 2013; Was, 2016; Zinkle and Was, 2013). Further being body-centred cubic (BCC), its distinct deformation properties, makes the effect of the loading rate significant. Plastic deformation in tungsten at low homologous temperatures is primarily controlled by the thermally activated movement of a/2<111> screw dislocations (Giannattasio et al., 2010; Gumbsch et al., 1998; Hirth and Lothe, 2016). The non-planar core arrangement of these dislocations results in a significant Peierls barrier, necessitating that plastic flow occurs via the nucleation and movement of kink pairs (Frost and Ashby, 1982; Hirth and Lothe, 2016). Thus, the flow stress is heavily influenced by temperature and strain rate. Increasing strain rate reduces the time available for thermal activation, requiring higher stresses to maintain plastic flow. This causes a transition towards increasingly brittle behaviour (Dümmer et al., 1998; Giannattasio and Roberts, 2007; Zheng et al., 2016). Multiple investigations have shown that tungsten shows significant strain-rate sensitivity, where higher loading rates lead to increased flow stresses, decreased ductility, and greater vulnerability to brittle failure (Dümmer et al., 1998; Giannattasio and Roberts, 2007; Miranda et al., 2026). Understanding how loading rate affects the fundamental dislocation mechanisms is thus crucial for predicting the deformation and failure behavior of tungsten in conditions relevant to its service.

Neutron irradiation of tungsten further complicates this behaviour by introducing a high density of defect obstacles through displacement damage and transmutation reactions, including vacancy clusters, dislocation loops, voids and Re/Os-rich precipitates (Das et al., 2019a, 2019b;

Dürrschnabel et al., 2021; Gilbert, 2010; Gilbert and Sublet, 2011; Lloyd et al., 2024, 2019; Rieth et al., 2013; Yi et al., 2013). These irradiation-induced defects impede dislocation motion, decrease the activation volume linked to plastic deformation, producing pronounced radiation hardening, a more localized plastic deformation, reduced ductility and a substantial increase in the brittle-to-ductile transition temperature (Abernethy, 2017; Abernethy et al., 2019; Bushby et al., 2012; Suchandrima Das et al., 2018; Hardie et al., 2015; Hasenhuetl et al., 2017; Makin and Sharp, 1965). Crucially, irradiation alters the defect-governed kinetics of thermally activated plastic flow without fundamentally changing the basic deformation mechanism (Abernethy et al., 2019; Das et al., 2019a). Given that strain-rate sensitivity and irradiation hardening arise from the interplay between mobile dislocations and thermally activated obstacles, a significant link may exist between the loading rate and the microstructure induced by irradiation. Understanding this correlation is especially needed during transient loading situations, where the time for thermal activation becomes progressively restricted.
Experimental studies on strain-rate sensitivity have utilized various approaches, including standard tensile tests, compression tests, split-Hopkinson pressure bar experiments, micro-pillar compression, micro-cantilever bending, and nanoindentation, among others (Dümmer et al., 1998; Zheng et al., 2016). As plasma-facing components are also irradiated in service, nanoindentation is particularly attractive, because it requires only small material volumes and is therefore suited both to radioactive specimens and to ion-irradiated layers only a few micrometres thick (Beck et al., 2017; Suchandrima Das et al., 2018; Das et al., 2019a; Pathak et al., 2017). Two nanoindentation routes are commonly used to probe rate-dependent plasticity. Strain-rate jump tests with a geometrically self-similar Berkovich indenter keep the representative strain approximately constant with depth, and so provide the strain-rate sensitivity and activation volume largely free of changes in deformation geometry; they have been applied to the neutron-irradiated tungsten studied here. (Abernethy et al., 2019; Maier et al., 2011). Strain-rate-controlled spherical indentation instead provides an indentation stress–strain curve from a single test, from the elastic regime through the elastic–plastic transition into plastic flow (Kalidindi and Pathak, 2008; Pathak et al., 2017). This makes it increasingly attractive for irradiated materials, in which the yield and early hardening response are of particular interest, but it comes at the cost of a continuously evolving deformation geometry.

In spherical indentation, the contact radius, representative indentation strain and plastic-zone size all increase with penetration depth (Maneiro and Rodríguez, 2004; Pathak et al., 2017; Spary et al., 2006). The measured response therefore reflects not only the intrinsic constitutive behaviour of the material, but also the evolving deformation length scale and the associated accumulation of geometrically necessary dislocations (GNDs) and strain-gradient hardening (Suchandrima Das et al., 2018; Fleck and Hutchinson, 1997; Gao et al., 1999; Kalidindi and Pathak, 2008; Liang and Pharr, 2022; Nye, 1953). In strain-rate-controlled tests, this coupling extends to the loading rate itself: the indentation strain rate is defined through the evolving contact geometry, rather than being imposed independently of the deformation field as in uniaxial testing (Kalidindi and Pathak, 2008; Oliver and Pharr, 2004). Changes in the measured load, hardness and pile-up can therefore reflect the combined effects of constitutive rate sensitivity, deformation length scale and strain-gradient hardening, rather than rate sensitivity alone (Hackett et al., 2024; Liang and Pharr, 2022). Unlike in Berkovich testing, this coupling cannot be avoided in strain-rate-controlled spherical indentation; it must instead be understood and accounted for when the results are interpreted. This is particularly important for neutron-irradiated materials, in which irradiation-induced defects restrict the growth of the plastic zone and promote highly localised deformation through dislocation–defect interactions (Byun and Hashimoto, 2006; Das et al., 2019a, 2019b; S. Das et al., 2018; Gussev et al., 2015; Onimus et al., 2004).

When combined with constitutive modelling, strain-rate dependent spherical nanoindentation thus provides a unique possibility for investigating how these coupled effects influence the measured load–displacement response, residual pile-up, and subsurface deformation, and for disentangling the respective contributions of loading rate and indentation depth. Previous studies have demonstrated that strain-gradient crystal plasticity finite element (CPFE) constitutive formulations incorporating thermally activated slip can reproduce load–displacement responses, pile-up morphologies, residual lattice-strain profiles, and GND-density distributions in unirradiated and ion-implanted materials (Das et al., 2020, 2019a, 2019b; Suchandrima Das et al., 2018). Beyond comparison with experiments, CPFE allows loading rate and indentation depth to be varied independently, enabling controlled numerical investigations of effects that remain coupled in strain-rate-controlled indentation experiments. While recent advancements have successfully utilized material-invariant parameters to predict deformation across varying loading histories and length scales without continuous recalibration (Ashraf and Castelluccio, 2023; Dindarlou and Castelluccio, 2022), previous investigations into irradiated microstructures have focused primarily on ion-implanted materials under quasi-static loading or on reproducing deformation at individual loading conditions. Consequently, applying a single physically informed constitutive framework to reproduce multiple experimental observables across different strain rates without further parameter recalibration, and subsequently disentangle the contributions of loading rate and deformation length scale, remains a significant challenge, particularly for neutron-irradiated materials.

Thus, in this study a combination of strain-rate controlled nanoindentation with CPFE modelling is used to study unirradiated and neutron irradiated tungsten. The continuous evolution of the contact radius, representative indentation strain, and plastic-zone size are used to examine how the deformation length scale interacts with material strain-rate sensitivity. Spherical nanoindentation tests are conducted at various loading rates. Atomic force microscopy (AFM) and high-resolution electron backscatter diffraction (HR-EBSD) are used to provide complementary measurements of residual surface morphology, lattice strain, lattice rotation, and GND density, enabling the simulated deformation fields to be assessed against multiple experimental observables (Britton and Hickey, 2018; Das et al., 2019a, 2019a; Jiang et al., 2013; Wilkinson et al., 2006). The experimental observations at one strain-rate are used to calibrate a strain-gradient CPFE framework which integrates thermally activated slip kinetics, irradiation hardening, and strain softening. Without further parameter changes, the framework is then assessed against the measured load–displacement responses, pile-up morphologies, residual lattice strain fields and GND distributions at the other loading rates. It is subsequently used to perform controlled numerical studies in which indentation depth and loading rate are varied independently, allowing the influence of strain-rate sensitivity, evolving deformation geometry and irradiation-induced defect interactions to be assessed systematically. Finally, these controlled simulations are used to interpret the experimental observations, distinguishing the contributions of loading rate and indentation depth to the measured response. The resulting analysis provides a mechanistic interpretation of strain-rate-controlled spherical indentation of irradiated materials, and a basis for developing constitutive descriptions of plasma-facing materials under transient loading.

## 2. Experimental Methodology

### 2.1 Materials and Sample Preparation

This research utilized two samples of commercially pure single-crystal tungsten (99.99%). The first was a disc-shaped specimen from Goodfellow, measuring 4 mm thick with a 7 mm diameter and an out-of-plane [001] orientation. The second sample was a 1 x 1 x 12 mm rod from Metal Crystals and Oxides Limited and specifically prepared for neutron irradiation. For the rod specimen, the normal to the (1 × 1) $mm^2$ cross-sectional face was parallel to [001], while the longitudinal axis lay in the plane perpendicular to [001]. The indented surfaces of both specimens were within 5° of (001) orientation. Both samples were polished using a multi-step process to achieve a smooth, stress-free surface for high-resolution analysis. This began with mechanical grinding using silicon carbide (SiC) paper in successive grits from 800 to 2500. Following this, the surfaces were polished with diamond pastes of 6 μm, 3 μm, and 1 μm, and the final mechanical step was performed with a colloidal silica suspension.

To remove any residual mechanical polishing damage, the samples were electropolished. The disc was electropolished in a 1% NaOH aqueous solution, while the rod was processed in a 0.5% NaOH aqueous solution. Both were done at room temperature. This final step ensured a high-quality surface suitable for techniques such as high-resolution electron backscatter diffraction (HR-EBSD) (details of the technique follow below, Section 2.5).

### 2.2 Neutron Irradiation

The tungsten specimens, originally machined into single crystal bars, were irradiated at elevated temperature (1173 K) for an extended duration of 208 full power days in the High Flux Reactor (HFR) at Petten, Netherlands, as part of the EXTREMAT-II campaign (2008) (Abernethy et al., 2019; Klimenkov et al., 2016; Lloyd et al., 2019). To ensure homogeneity of neutron exposure, the specimens—fabricated in the form of matchstick geometries were sequentially positioned at two irradiation sites within the reactor core, undergoing 148 and 60 days of exposure respectively. The local neutron fluxes at these positions were $6.8 \times 10^{18}$ m$^{-2}$ s$^{-1}$ and $6.6 \times 10^{18}$ m$^{-2}$ s$^{-1}$, with corresponding fast neutron flux components ($E > 0.1$ MeV) of $3.2 \times 10^{18}$ m$^{-2}$ s$^{-1}$ at both sites (Lloyd et al., 2024). A study using FISPACT-II simulations, estimated that a damage level of 1.67 dpa would result from irradiation at a dose rate of $9 \times 10^{-8}$ dpa·s$^{-1}$, assuming a displacement threshold energy of 55 eV (Abernethy et al., 2019). The predicted alloy composition was W – 1.4 at.% Re – 0.1 at.% Os – 0.02 at.% Ta in good agreement with FISPACT modelling and EDX measurements, which predicted W1.4Re-0.1Os (Klimenkov et al., 2016). However, the neutron energy spectrum of the sample was reduced due to the presence of a neighbouring specimen at both positions, which acted as a strong absorber of thermal neutrons, as previously reported (Abernethy et al., 2019). Post-irradiation characterization using Energy Dispersive X-ray Spectroscopy (EDX) and Atom Probe Tomography (APT) (Gault et al., 2021) revealed a composition of W – 1.20 ± 0.11 at.% Re – 0.11 ± 0.05 at.% Os – 0.03 ± 0.01 at.% Ta (Lloyd et al., 2019).

### 2.3 Nanoindentation Procedure

Spherical nanoindentation was performed using an Agilent G200 nanoindenter in continuous stiffness measurement (CSM) mode, with a displacement oscillation amplitude of 2 nm at 45 Hz. A spherical diamond tip with a nominal radius of 5 μm was calibrated against fused silica using the spherical tip area-function approach of (Leitner et al., 2018). The calibrated contact area was expressed as a function of contact depth as

$$A_c(h_c) = B_0 h_c^{B_1} - B_2 h_c^2, \qquad (1)$$

where $B_0$, $B_1$, and $B_2$are the geometric constants obtained from tip calibration.

The contact depth $h_c$ was calculated from the measured indentation displacement, load, and contact stiffness according to

$$h_c = h - \epsilon_{\mathrm{OP}} \frac{P}{S}, \qquad (2)$$

where $h$ is the displacement into the specimen surface, $P$ is the indentation load, $S$ is the contact stiffness measured using CSM, and $\epsilon_{\mathrm{OP}}$ is the indenter geometric constant. Thus, the correction from the elastic displacement outside the contact is included in the determination of $h_c$.

The contact radius $a_c$ was then obtained from the projected contact area according to

$$a_c = \sqrt{\frac{A_c}{\pi}}. \qquad (3)$$

The indentation strain was defined throughout the experiment using the indentation-strain measure proposed by Kalidindi and Pathak:

$$\varepsilon_i = \frac{4h}{3\pi a_c}, \qquad (4)$$

where $a_c$ is the instantaneous contact radius. Here, $\varepsilon_i$ denotes the physical indentation strain used to describe the spherical indentation response. The strain rates specified in this work refer to the prescribed rate of this same indentation-strain measure, i.e. $\dot{\varepsilon}_i = d\varepsilon_i/dt$.

**Initial loading stage**

At the beginning of an indentation, the strain-rate feedback channel cannot be used immediately because a finite penetration depth is required to establish the contact geometry. Therefore, the indentation was initially performed under displacement control at a constant displacement rate of 0.5 nm $s^{-1}$.

During this initial stage, the contact radius was obtained using the ideal spherical geometry,

$$a_c = \sqrt{2Rh_c - h_c^2}, \qquad (5)$$

where $R$ is the nominal radius of the spherical indenter. This geometrical relation was used only during the initial loading stage, instead of the calibrated area function in Eq. (3), to establish the indentation strain before the strain-rate feedback became active.

Combining Eqs. (4) and (5) gives the relationship between contact depth and indentation strain for an ideal spherical indenter:

$$h_c = \frac{18\pi^2 \varepsilon_i^2 R}{16 + 9\pi^2 \varepsilon_i^2}. \qquad (6)$$

For a prescribed indentation strain rate $\dot{\varepsilon}_i$, differentiation of Eq. (6) gives

$$\dot{h}_c = \frac{576\pi^2 R \varepsilon_i \dot{\varepsilon}_i}{(16 + 9\pi^2 \varepsilon_i^2)^2}. \qquad (7)$$

Eqs. (6) and (7) are idealised closed-form relations for a perfectly spherical indenter, obtained by combining Eqs. (4) and (5) under the simplifying assumption $h \approx h_c$, i.e. neglecting the elastic displacement outside the contact. In the experiments, the indentation strain was always evaluated from Eq. (4) using the measured displacement $h$ and the contact radius obtained from the corrected contact depth $h_c$ (Eq. 2), via Eq. (5) during the initial stage and the calibrated area function (Eq. 3) thereafter (as discussed below). Eq. (7) is used in Section 4.1 to convert the target indentation strain rates into indenter displacement rates for the CPFE simulations, in which the indenter is an ideal sphere of radius R.

An initial indentation strain of $\varepsilon_i = 0.02$ was reached using the constant displacement rate of 0.5 nm $s^{-1}$. Once this initial strain was reached, the control mode was changed from displacement control to strain-rate control.

**Strain-rate-controlled stage**

Following the transition at $\varepsilon_i = 0.02$, the calibrated spherical tip area function was used to determine the contact area from the measured contact depth. The instantaneous contact radius was therefore calculated using Eq. (3), and the indentation strain was continuously evaluated from Eq. (4).

The time derivative of the indentation-strain channel,

$$\dot{\varepsilon}_i = \frac{d\varepsilon_i}{dt}, \qquad (8)$$

was then used as the feedback quantity for the proportional-integral-derivative (PID) control system. The PID controller adjusted the indenter displacement so that the measured indentation strain rate approached the user-defined target value. Thus, although the physical motion of the indenter is displacement, the principal control variable during the main portion of the test was the indentation strain rate.

The indentation was continued under strain-rate control until a total indentation strain of $\varepsilon_i = 0.1$ was reached. Three target indentation strain rates were investigated:

$$3.2 \times 10^{-5}\ \mathrm{s}^{-1},\ 3.2 \times 10^{-4}\ \mathrm{s}^{-1},\ 3.2 \times 10^{-3}\ \mathrm{s}^{-1}.$$

The PID parameters were tuned to minimise variation in the achieved strain rate; the parameters used were $P = 0.5$, $I = 0.001$, and $D = 0.015$. In absence of the time records of the achieved strain rate for the individual tests, the deviation from the target strain rate cannot be quantified. The strain rates are therefore referred to as nominal strain rates throughout.

The two stages should therefore be distinguished as follows: the initial stage used constant displacement-rate control to establish a finite penetration and an initial indentation strain of 0.02, with the ideal spherical geometry used to determine the contact radius; the subsequent stage used the calibrated tip area function and feedback control of the indentation-strain rate until $\varepsilon_i = 0.1$. The definition of indentation strain remained the same in both stages.

All nanoindentation tests were performed at room temperature. Nine indents were performed for each material and nominal strain-rate condition. Owing to the stochastic pop-ins in the load–displacement response, a representative single indent is presented for each condition. No thermal drift correction was applied. Given the considerably longer duration of the slowest tests (approximately 2500 s for the strain-rate-controlled segment at $3.2 \times 10^{-5}$ s$^{-1}$, compared with approximately 25 s at $3.2 \times 10^{-3}$ s$^{-1}$), thermal drift may contribute to the measured displacement at the lowest strain rate (see Section 3).

### 2.4 Atomic force microscopy (AFM)

For analysing the surface morphology of the samples, a Dimension 3100 (Nanoscope IV) atomic force microscope (AFM) from Veeco/Bruker was used. The instrument was operated in contact mode to examine the topography near the nano indents in unirradiated and neutron-irradiated single-crystal tungsten samples. A silicon tip with a nominal radius of 8 nm was employed for all measurements. The raw AFM data was processed with Gwyddion, an open-source software (Nečas and Klapetek, 2012), to generate topographical maps. For scans larger than 10 μm in diameter, a background curvature was present due to the cantilever's arc-like motion. This artifact was mathematically removed by applying a quadratic fit to the data, ensuring the preservation of the actual surface topography.

### 2.5 High-Resolution Electron Backscatter Diffraction (HR-EBSD)

Elastic lattice strain and lattice rotation fields around the indents were measured by high-resolution electron backscatter diffraction (HR-EBSD) (Britton and Wilkinson, 2012, 2011; Wilkinson, 1996; Wilkinson et al., 2006). Maps were acquired in a Tescan MIRA3 XHM scanning electron microscope with an Oxford Instruments Symmetry S2 detector and AZtec 5.0, at 30 kV and 15 nA, using a step size of 50 nm and 1×1 binning. Patterns were analysed by cross-correlation in the XEBSD code, relative to a reference pattern from a nominally strain-free region of the same grain, to obtain the lattice rotation tensor and the deviatoric elastic strain tensor. GND densities were computed from the lattice curvature over the 12 a/2⟨111⟩{110} and 12 a/2⟨111⟩{112} slip systems using L2 minimisation (Das et al., 2020). As the step size defines the length scale over which lattice curvature is evaluated, only dislocation content producing measurable orientation gradients over 50 nm contributes to the computed GND density; this step size was chosen to balance spatial resolution against measurement noise (Jiang et al., 2013).

## 3. Experimental interpretation and motivation for predictive CPFE modelling

The nanoindentation experiments show systematic differences between the nominal strain-rate conditions in both unirradiated and neutron-irradiated tungsten, with a reduction in indentation load **Fig. 1** and surface pile-up (**Fig. 2**) from the fastest to the slowest nominal strain rate. However, interpreting these trends as intrinsic strain-rate sensitivity is complicated by the strain-rate feedback protocol used during testing (as detailed in Section 2.3).

Although all tests targeted the same final indentation strain, experiments performed at different nominal strain rates reached significantly different final penetration depths. For the unirradiated material, the maximum depths reached were approximately 528 nm, 430 nm, and 360 nm for the fast, intermediate, and slow strain-rate conditions, respectively. The irradiated material exhibited a similar depth variation, reaching approximately 538 nm, 450 nm, and 336 nm under the corresponding loading conditions. Under the feedback protocol, such differences could arise from intrinsic rate sensitivity of the material response, or from instrumental and control-related effects, such as thermal drift over the considerably longer duration of the slower tests or lag in the feedback control. These contributions cannot be separated from the experiments alone. The experimentally measured load–displacement response, pile-up morphology and lattice deformation are therefore treated as reflecting the combined influence of the nominal loading rate and the different indentation depths, and hence plastic-zone sizes, attained in each test.

To decouple these variables, the strain-rate-sensitive crystal plasticity finite element (CPFE) framework serves as a controlled computational platform. Unlike the experimental protocol, which controls the target indentation strain rate through feedback from the evolving indentation response, the CPFE simulations prescribe the indenter displacement history directly as a boundary condition. This approach allows the displacement rate and final indentation depth to be varied independently. Thermally activated slip, strain localization, irradiation-induced hardening, and strain-gradient evolution remain fully coupled within the material constitutive formulation, but the computational framework removes the additional geometric coupling imposed by the experimental feedback system. In the following sections, the strain-rate-sensitive CPFE formulation is described in detail, followed by a systematic comparison between experimental observations and model predictions.

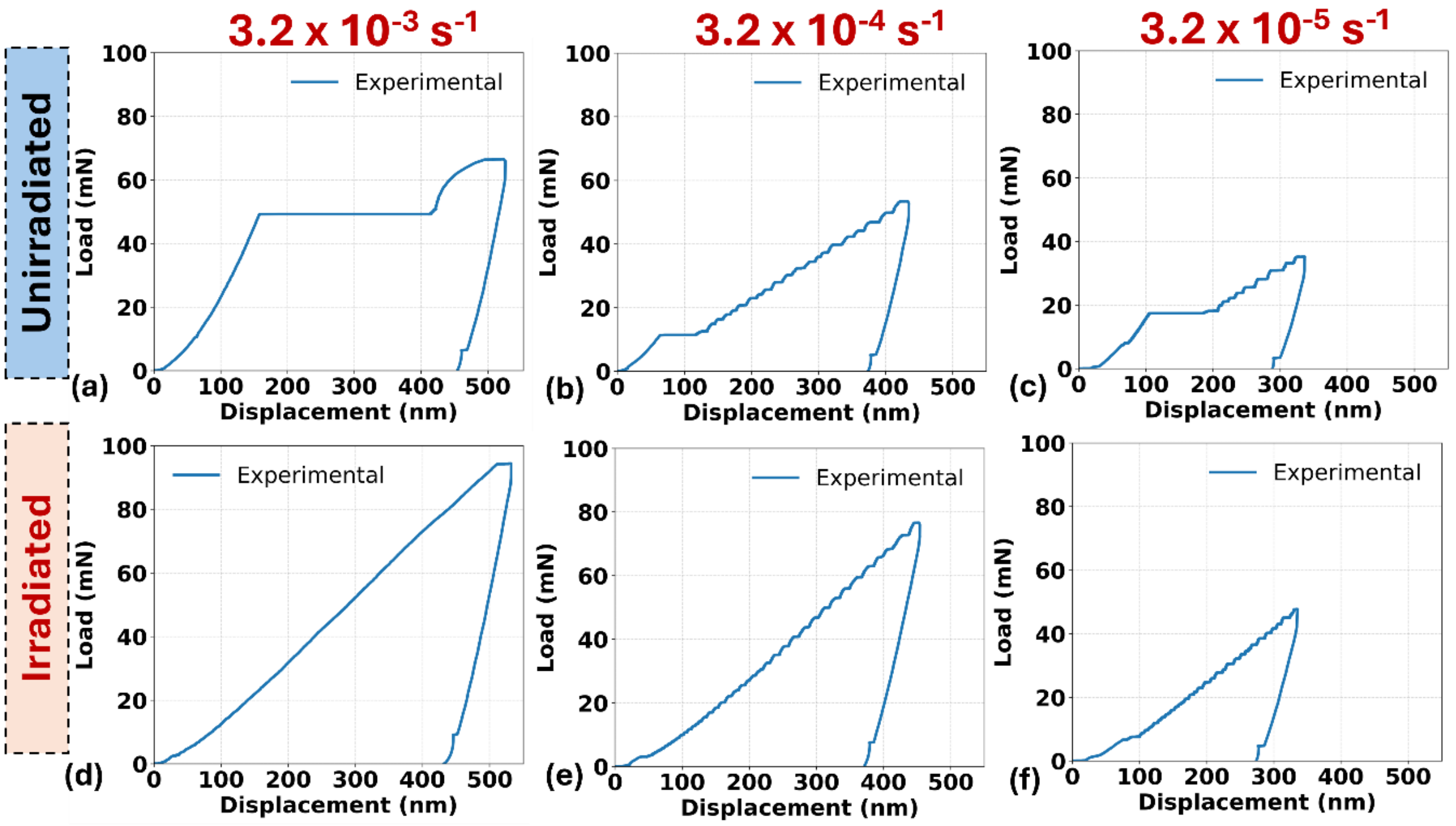

**Fig. 1**. Load-displacement curves as measured by nano-indentation for unirradiated and neutron irradiated single crystal tungsten for strain rates $3.2 \times 10^{-3}$ s$^{-1}$, $3.2 \times 10^{-4}$ s$^{-1}$, and $3.2 \times 10^{-5}$ s$^{-1}$.

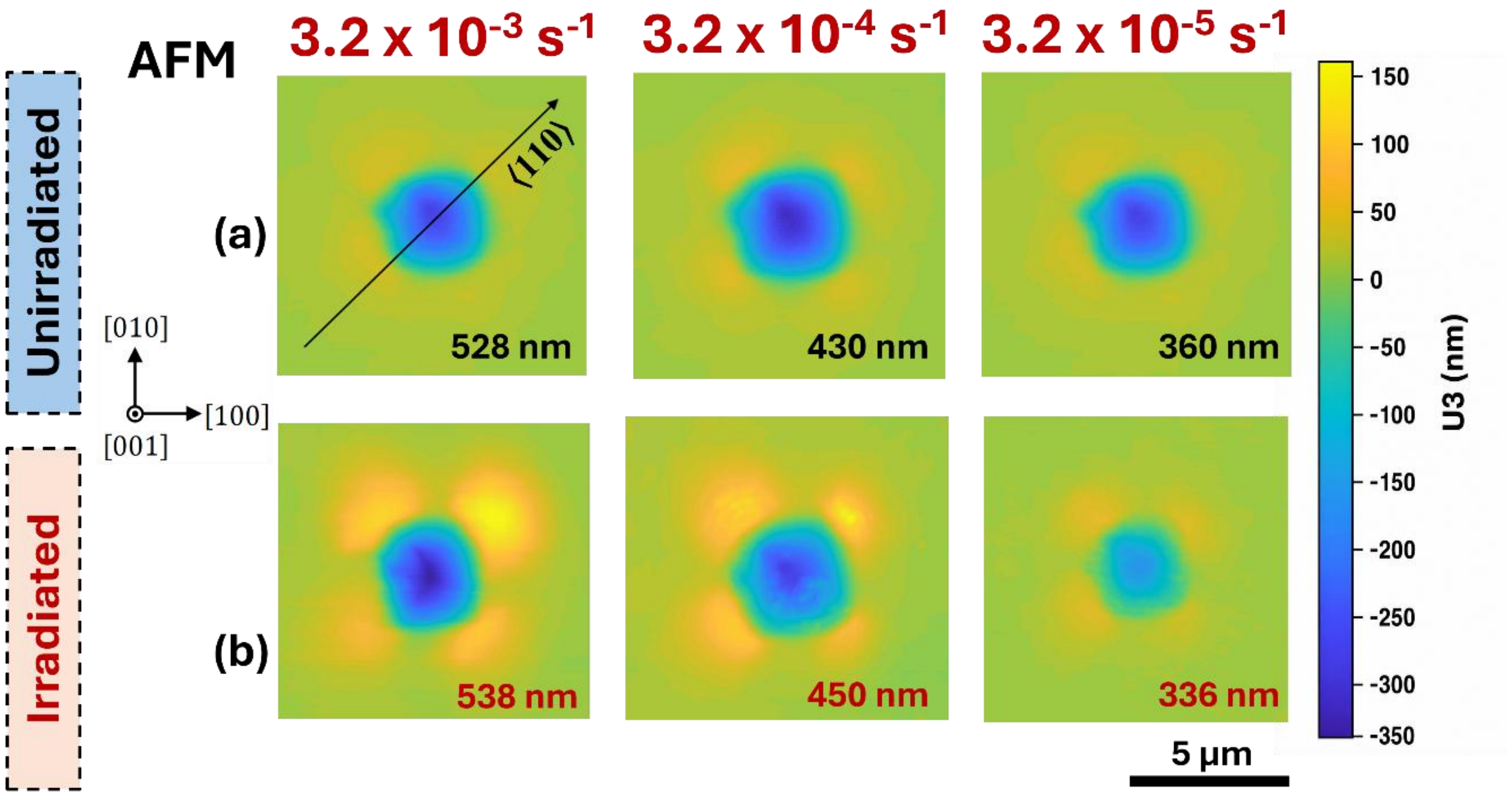

**Fig. 2.** Surface morphologies (height profile) of nano-indents in (a) pure tungsten and in (b) neutron irradiated tungsten measured by AFM after indentation for strain rates 3.2 x $10^{-3}$ s$^{-1}$, 3.2 x $10^{-4}$ s-1, and 3.2 x $10^{-5}$ s$^{-1}$. The maximum depth reached for each strain rate is annotated in the figure.

## 4. CPFE formulation

The nano-indentation experiments were simulated using a strain-gradient crystal plasticity model where dislocation slip was restricted to occur in slip directions compatible with the crystallography. The finite element software Abaqus 2025 (Dassault Systèmes, Providence, RI, USA), was used to simulate the indentation process. The crystal plasticity model was implemented in an Abaqus user material subroutine (UMAT). The model was constructed as having <001> crystallographic orientation, similar to the physical sample.

### 4.1 The CPFE model

A three-dimensional finite element model was developed to simulate the nanoindentation response of tungsten. The model consists of a deformable cubic sample block of dimensions 20 × 20 × 20 $\mu m^3$ and a rigid spherical indenter with a radius of 5 µm, as shown schematically in **Fig. S.1**. Owing to the crystallographic symmetry associated with the <001> grain orientation and the axisymmetric nature of spherical indentation, only one quarter of the experimental configuration was modelled. Corresponding symmetry boundary conditions were imposed on the XZ and YZ planes. The top surface of the sample was traction-free, while the remaining external surfaces were constrained to prevent rigid body motion, consistent with the experimental boundary conditions. The indenter was modelled as a discrete rigid body, and a frictionless hard contact was prescribed between the indenter and the sample surface. This assumption is justified by prior studies demonstrating that the load–displacement response during spherical indentation is largely insensitive to the friction coefficient at the indenter–sample interface (Wang et al., 2004). The deformable sample block was discretized using second-order, 20-node three-dimensional hexahedral elements with reduced integration (C3D20R) (Das et al., 2020; Demir et al., 2024). The simulations were performed with geometric nonlinearity (NLGEOM in Abaqus), so that the large plastic deformations and lattice distortions beneath the indenter, and the resulting pile-up, are captured within a finite-deformation framework with the deformation gradient decomposed multiplicatively into elastic and plastic parts. A spatially refined mesh was employed in the vicinity of the indentation site to accurately capture the high stress and strain gradients beneath the indenter, with element sizes ranging from approximately 0.16 µm near the contact region to 2 µm in the far field. The final mesh comprised approximately 19683 elements, providing a balance between numerical accuracy and computational efficiency. Mesh refinement and convergence were carefully assessed, as numerical accuracy in indentation simulations is highly sensitive to element size near the contact zone (Choi et al., 2008; Izadpanah Najmabad et al., 2024; Weiss et al., 2024; Weiss and Knezevic, 2024). Convergence was deemed satisfactory when further mesh refinement resulted in changes of 3–5% or less in the predicted response. In the present study, a minimum element size of 160 nm with an applied edge bias ranging from 0.16 to 2 μm was found to be sufficient. The predicted maximum pile-up height stabilized for mesh sizes below approximately 0.55 μm, with the selected mesh size of 0.16 μm providing a mesh-independent, computationally efficient solution (**Fig. S.2**).

Indentation was simulated under displacement control: the indenter was driven to the experimentally achieved depth for each condition and then unloaded, consistent with the experimental protocol. The experiments prescribed the indentation strain rate through feedback control. However, the CPFE simulations require the indenter displacement rate to be prescribed directly. Thus, each target indentation strain rate was converted to a displacement rate using Eq. (7) with R = 5 μm. As the strain beneath a spherical indenter is spatially heterogeneous and evolves during loading, a single representative strain, ε_rep = 0.008, was used in this conversion (when strain contours are plotted within fixed limits of ±0.008, it resolved strain localisation beneath the indenter without saturation as shown in Section 4). For a fixed ε_rep, Eq. (7) makes the displacement rate proportional to the target strain rate, so the prescribed displacement rates retain the ten-fold ratios of the experimental strain rates. Any difference between ε_rep and the effective strain of the experiments scales all displacement rates by a common factor, whose effect on the flow stress is absorbed when the model is calibrated at the reference strain rate; this is shown in detail in **Appendix D**. The prescribed displacement rate is therefore not a local strain-rate boundary condition. The local strain and strain-rate fields develop from the indentation kinematics, the evolving contact geometry and the constitutive response, and are spatially heterogeneous, particularly beneath the indenter, where pile-up and localised plastic deformation produce strong gradients. Simulated indentation loads were corrected for indenter compliance using an effective elastic modulus (Das et al., 2020). For both material states, the model was calibrated against the experimental load–displacement response and pile-up at the highest nominal strain rate ($3.2 \times 10^{-3}$ $s^{-1}$). Simulations at the two lower strain rates were then performed by changing only the prescribed displacement rate and final depth, with all other aspects of the model unchanged (Section 4.3).

### 4.2 UMAT formalism

#### 4.2.1 Pure Tungsten

The crystal plasticity formulation employed in this work is implemented within an Abaqus user material subroutine (UMAT), as reported by (Das et al., 2020, 2019a; Suchandrima Das et al., 2018). Full details of the numerical implementation are provided elsewhere (Suchandrima Das et al., 2018; Dunne et al., 2007). Briefly, the constitutive description adopts a physically based slip law that accounts for the thermally activated glide of dislocations through a field of pinning obstacles.

The crystallographic slip rate, $\dot{\gamma}$, on a given slip system is governed by a physically based constitutive law that accounts for the thermally activated glide of dislocations through a field of pinning obstacles. In current CPFE framework, the slip rate on slip system $\lambda$ is expressed as

$$\dot{\gamma} = \rho_m (b^{\lambda})^2 \nu \exp\left(-\frac{\Delta F}{kT}\right)\sinh\left(\frac{(|\tau^{\lambda}|-\tau_c)\Delta V}{kT}\right)\mathrm{sgn}(\tau^{\lambda}) \quad (9)$$

where $\rho_m$ is the density of mobile dislocations, $\nu$ is the attempt frequency, $b$ is the magnitude of the Burgers vector, $\Delta F$ is the activation energy, $k$ is the Boltzmann constant, and $T$ is the absolute temperature. The resolved shear stress on slip system $\lambda$ is denoted by $\tau^{\lambda}$, while $\tau_c$ represents the critical resolved shear stress (CRSS).

The activation volume $\Delta V$ depends on the average spacing between pinning dislocations, $l$, which is estimated as

$$l = \frac{1}{\sqrt{\Psi(\rho_{\mathrm{SSD}})}}, \quad (10)$$

where $\Psi$ represents the probability of pinning determined by the density of statistically stored dislocations, $\rho_{\mathrm{SSD}}$. Within the present CPFE framework, $\rho_{\mathrm{SSD}}$ has not evolved explicitly during deformation. This assumption is motivated by the strain-gradient-dominated nature of nanoindentation in single-crystal tungsten, where geometrically necessary dislocations (GNDs) are known to dominate the dislocation content. Previous studies combining Laue diffraction measurements with CPFE simulations have shown that GND densities beneath nanoindents in pure tungsten can reach values of the order of $10^{17}\ \mathrm{m}^{-2}$ (Suchandrima Das et al., 2018). In comparison, the contribution of statistically stored dislocations is expected to be negligible. Accordingly, $\rho_{\mathrm{SSD}}$ and the mobile dislocation density $\rho_m$ are assumed to be significantly smaller than the GND density and are taken to be of the order of $10^{10}\ \mathrm{m}^{-2}$. This treatment ensures that the strain-gradient-controlled response characteristic of nanoindentation is appropriately captured. It is recognised that plastic deformation, particularly in the presence of irradiation-induced obstacles, can generate and redistribute dislocations and therefore alter both the SSD and mobile dislocation populations. A fully evolving treatment would require constitutive laws describing dislocation multiplication, recovery and the coupling between the SSD, mobile and GND populations, together with additional parameters that cannot be independently constrained by the experimental observables available in the present study. These quantities were held fixed to constrain the number of independently calibrated constitutive parameters, as the available experimental observables do not provide independent constraints on their evolution. The fixed-density treatment is thus a reduced-order modelling assumption adopted to isolate the effects of the calibrated constitutive parameters. In this study, CPFE predictions are performed by considering the contributions from the 12 $a/2\langle 111\rangle\,\{110\}$ and 12 $a/2\langle 111\rangle\,\{112\}$ slip systems (Total of 24) which are the primary slip families experimentally observed to operate under room-temperature (Marichal et al., 2013; Weinberger et al., 2013).

The strain-rate sensitivity in the present model is governed by the activation energy barrier ΔF, which is used in the slip law (Eq. 9) (Zheng et al., 2016). Following (Gumbsch et al. (1998)), an activation energy of $\boldsymbol{\Delta F = 3.4559 \times 10^{-20}}$ **J** (0.2 eV) was adopted, corresponding to the reported activation energy for thermally activated dislocation motion in [001]-oriented tungsten. The same value was used for both unirradiated and irradiated tungsten, while the effects of irradiation were represented through the irradiation-specific hardening and obstacle contributions in the constitutive framework. This value was retained during the subsequent strain-rate simulations after calibration against the reference experimental condition. The values of the material parameters used here are provided in **Appendix C**. Isotropic elasticity is assumed, and the elastic strains are assumed to be small; this assumption applies to the elastic deformation only, while large plastic deformations and rotations are accounted for through the finite-deformation (NLGEOM) analysis. Isotropic elasticity adopted for the elastic response of tungsten is justified by its exceptionally low elastic anisotropy; experimental measurements of single-crystal tungsten give a Zener anisotropy ratio close to unity (Featherston and Neighbours, 1963a). Anisotropy in plastic deformation, is explicitly captured through the crystallographic slip systems in the CPFE formulation. Also, for simplicity, all parameters on the RHS in Eq. (9) were kept unchanged, other than $\tau_c$ (considered the same for all slip-systems) and the independent variable $\tau^\lambda$.

Crystallographic slip occurs when the resolved shear stress $\tau^\lambda$ exceeds the critical resolved shear stress $\tau_c$, which evolves according to

$$\tau_c = \tau_{c_0} + C'Gb\sqrt{\rho_{\mathrm{GND}}} \qquad (11)$$

where $\tau_{c_0}$ is the initial CRSS of the unirradiated tungsten sample. The second term represents strain hardening due to the accumulation of geometrically necessary dislocations generated to accommodate lattice curvature. Here, $C'$ is a geometric hardening factor that accounts for the fraction of generated dislocations contributing to hardening, $G$ is the shear modulus of tungsten, and $\rho_{\mathrm{GND}}$ is the total GND density summed over all active slip systems. The GND density is computed from the curl of the plastic deformation gradient tensor, following the curl formulation as stated in (Arsenlis and Parks, 1999; Suchandrima Das et al., 2018). Thus, overall, there are two tuning parameters in the UMAT formulation for pure tungsten $\tau_{c_0}$ and $C'$.

### 4.2.2 Neutron Irradiated Tungsten

To accurately represent the influence of irradiation-modified material on the indentation response, the tungsten block was assigned material properties corresponding to the irradiated condition throughout its volume. This is appropriate because neutron irradiation produces damage throughout the bulk of the specimen. The UMAT applied to the pure tungsten as described above is extended to irradiated tungsten. The parameters tuned for pure tungsten, i.e. $\tau_{c0}$ and $C'$ are kept unchanged and carried over in the UMAT formalism for the irradiated material.

In the irradiated sample, the model incorporates strain softening to represent the progressive clearance of defect obstacles by gliding dislocations (Xu et al., 2017). Initially, irradiation-induced defects act as strong barriers, increasing the resistance to plastic flow. As deformation proceeds, repeated interactions weaken and remove these defects, causing a localized reduction in flow resistance. This promotes the preferential accumulation of plastic strain in softened regions. In BCC metals like tungsten, the high initial irradiation resistance heavily restricts the cross-slip of screw dislocations, forcing successive dislocations to glide continuously on the same atomic planes and leading to the well-known phenomenon of slip channel formation (Das et al., 2019a; Fournier et al., 2009; Long et al., 2016; Luft, 1991; Onimus et al., 2004). To represent this mechanism within the CPFE framework, the critical resolved shear stress $\tau_c$ is allowed to evolve through an additional irradiation-induced hardening contribution that undergoes strain-dependent softening. The accumulated crystallographic slip at the end of each time increment $\Delta t$ is first evaluated as

$$\gamma^{t+\Delta t} = \gamma^{t} + \sum_{\lambda=1}^{n} \dot{\gamma}^{\lambda}\, \Delta t, \quad (12)$$

where $\dot{\gamma}^{\lambda}$denotes the slip rate on slip system $\lambda$, and the summation is taken over all active slip systems. This accumulated slip serves as a measure of the local plastic activity experienced by the material point. The irradiation-induced contribution to the CRSS, denoted $\tau_H$, is then reduced at the end of each increment according to an exponential softening law,

$$\tau_H^{t+\Delta t} = \tau_H^{0} \exp\left(-\frac{\gamma^{t+\Delta t}}{\gamma_s}\right), \qquad (13)$$

where $\tau_H^0$ represents the initial irradiation-induced hardening and $\gamma_s$ characterizes the accumulated slip at which the irradiation-induced resistance begins to saturate and transition into softening. The exponential form is adopted as a phenomenological representation of progressive irradiation-defect removal during plastic deformation. Specifically, the formulation assumes that the rate of defect removal is proportional to the population of irradiation-induced obstacles that remains available for interaction with dislocations. This first-order removal assumption gives an exponential decay of the surviving obstacle population with accumulated slip and, consequently, an exponential reduction in the associated irradiation-induced resistance.

This assumption is motivated by prior observations of defect removal and deformation-induced microstructural evolution in irradiated tungsten, where dislocation–defect interactions can progressively remove or weaken irradiation-induced obstacles as plastic slip accumulates (Das et al., 2019a). Accordingly, the evolution of the irradiation-induced hardening contribution $\tau_H$ with accumulated crystallographic slip $\gamma$ may be expressed as,

$$\frac{\partial \tau_H}{\partial \gamma}\Big|_{t+\Delta t} = -\frac{\tau_H}{\gamma_s}, \quad (14)$$

where $\gamma_s$ is a characteristic softening parameter controlling the rate at which irradiation-induced obstacles are weakened by plastic deformation. Integration of Eq. (14) over a time increment $\Delta t$ yields the exponential decay relation given in Eq. (13). Within this formulation, the total critical resolved shear stress governing slip in the irradiated material is composed of the base contribution associated with the unirradiated crystal and an additional irradiation-induced term that evolves according to Eqs. (11) and (13). The resulting CRSS is therefore written as

$$\tau_c = \tau_{c_0} + C'Gb\sqrt{\rho_{\text{GND}}} + \tau_H, \qquad (15)$$

where $\tau_{c_0}$ denotes the initial CRSS of the unirradiated material, $C'$ is the hardening coefficient, $G$ is the shear modulus, $b$ is the Burgers vector magnitude, and $\rho_{\text{GND}}$ is the total geometrically necessary dislocation density. The resulting spatial variation in $\tau_c$ naturally enables localized strain softening and directed plastic flow to emerge within the simulations, without the need for externally imposed localization criteria. $\tau_H^0$ is estimated analytically based on TEM information on the defect microstructure in the irradiated material and $\gamma_s$ is tuned to the experimental data.

#### 4.2.3 Determining $\tau_H^0$

The irradiation-induced defect hardening term $\tau_H^0$ was derived from experimentally observed defect populations obtained from STEM-HAADF characterization of neutron-irradiated single-crystal tungsten (Lloyd et al., 2022). The microstructure exhibited a varied distribution of dislocation loops, Re/Os-decorated voids, and Re/Os-rich precipitates resulting from irradiation damage and transmutation-driven solute segregation. The obtained defect-size ranges and documented number densities were later incorporated into the strengthening model to assess their separate and combined effects on dislocation pinning and the ensuing macroscopic hardening response. The approach is illustrated schematically in **Fig. 3**. Each defect population is distributed uniformly within the representative volume and projected onto a representative slip plane, the resulting mean obstacle spacing determines the stress required for a gliding dislocation to overcome that population, and the contributions of the three populations are combined to give $\tau_H^0$ (Section 4.2.3.4).

For each defect type, the following steps were adopted to estimate the areal density for a given size class i.e. the upper limit of the number of defects of size class $i$ per unit area that an advancing dislocation segment may intersect within a single layer. Transmission electron microscopy (TEM) observations report a total number density (*N)* of approximately in $m^{-3}$ for each defect type and provide a size range. *N* multiplied by the simulated volume (with dimensions 20 x 20 x 20 $\mu m^3$) then gave $N_v$, the total number of defects of that type that may be expected within the simulated volume. Since the detailed distribution of defect sizes within this population is not known a normal distribution is assumed to allocate the defects across chosen representative sizes resulting in size-resolved defect counts of $N_i$ for each size class *i.* For each size class, a uniform spatial distribution of defects was then assumed within the simulated volume. The cube was discretised along the Z axis into layers, each with a thickness equal to the corresponding defect diameter. Thus, the numbers of layers of defects, for a given size-class, equals the edge length of the cube (20 μm), divided by the diameter of the considered defect size. Here, we assume that the defects are perfectly circular. Thus, the number of defects per layer represents the upper limit of the number of defects that a dislocation may encounter as it is passing through this layer. This is denoted as $L_i$. Thus, $L_i$ corresponds to the number of defects of a given diameter in a plane of area (20 x 20 μm$^2$) (Table 1). To obtain the corresponding areal loop density ($\rho_i$), each value $L_i$ was normalized by the planar area of the layer. Since every layer spans the full cross-section of the simulation cell, the in-plane area is constant and equal to 20 μm x 20 μm = 400 μm$^2$. The areal density for defect size class $i$ is therefore given by

$$\rho_i = \frac{L_i}{400\ \mu m^2}, \quad (16)$$

### 4.2.3.1 Contribution from loops

The defect hardening contribution from dislocation loops was quantified by explicitly accounting for the finite population and size distribution of irradiation-induced dislocation loops within the representative simulation volume. Transmission electron microscopy (TEM) observations report $N$ of approximately $10^{20}$ m$^{-3}$ in the size range of 20–70 nm (Lloyd et al., 2022). This gave $N_v$ of ~ 8 x $10^5$ irradiation-induced loops. Representative loop diameters selected in this study are 20, 30, 40, 50, 60, and 70 nm. This procedure resulted in the size-resolved loop counts $N_i$ , $L_i$, and $\rho_i$ listed in Table 1.

Now we assume that the gliding dislocation cuts directly through the loop, and that the principal energy barrier to overcome this obstacle is the energetic cost of forming jogs. For an edge-on interaction, cutting through a loop produces a single jog of length equal to the Burgers vector magnitude $b$. The energy required to form this jog is given by

$$E_{jog} = \frac{Gb^3}{2}, \quad (17)$$

where $G$ is the shear modulus. The mechanical work performed in advancing a dislocation segment of length $L$ by a distance $b$ under an applied shear stress $\tau$ is

$$W = \tau\, b^2\, L, \quad (18)$$

Equating the jog-formation energy with the mechanical work at the critical stress $\tau_0$ yields

$$\tau_i = \frac{Gb}{2L}, \quad (19)$$

Following the standard forest-hardening framework, the obstacle spacing is related to the cutting density through

$$L = \frac{1}{\sqrt{\rho_i}}, \quad (20)$$

which leads to

$$\tau_i = \frac{Gb}{2}\sqrt{\rho_i}, \quad (21)$$

Equation (21) therefore provides the critical resolved shear stress for edge-on loop encounters, where a single jog is created. In contrast, for face-on interactions the dislocation must generate two jogs simultaneously, doubling the energetic cost. In this case, the strengthening contribution becomes

$$\tau_i = Gb\sqrt{\rho_i}, \quad (22)$$

To account for the general case in which both interaction geometries occur, the expression may be written in the unified form

$$\tau_i = m\,\frac{Gb}{2}\sqrt{\rho_i} \quad (23)$$

where $1 \leq m \leq 2$ is a geometry factor representing the average number of jogs formed per interaction, ranging from m = 1 for edge-on to m = 2 for face-on encounters. As the relative frequency of the two interaction geometries is not known for the present microstructure, the mean value m = 1.5 is adopted, corresponding to no preferred interaction geometry. The resulting stress $\tau_i$ for each loop size class is listed in Table 1. The planar densities $\rho_i$ represent upper limits on the number of loops a dislocation may intersect, as they assume that every loop within a layer lies on the slip plane.

This process is repeated across all loop sizes. To account for the contributions from the different loop-diameter classes, the individual hardening stresses were combined using the root-sum-square (RSS) superposition rule. This approach accounts for the contributions of the distinct loop populations without directly adding their stress increments linearly. The total loop-induced hardening stress is therefore expressed as

$$\tau_{\text{loop}} = \sqrt{\sum_{i=1}^{6} \tau_i^2} = \boxed{70.18 \text{ MPa}}$$

| **Loop Diameter (nm)** | **Layers in 20 µm** | **Total Loops ($N_i$)** | **Loops per Plane ($L_i$)** | **$\rho_i = L_i / 400$ (loops/µm²)** | **$\tau_i$ (MPa)** |
|---|---|---|---|---|---|
| **20** | **1000** | **55,355** | **55.36** | **0.1384** | **12.31** |
| **30** | **666.67** | **134,646** | **201.97** | **0.505** | **23.51** |
| **40** | **500** | **209,998** | **420.00** | **1.050** | **33.9** |
| **50** | **400** | **209,998** | **525.00** | **1.3125** | **37.91** |
| **60** | **333.33** | **134,646** | **403.94** | **1.00985** | **33.25** |
| **70** | **285.71** | **55,355** | **193.74** | **0.48435** | **23.02** |
| **RSS TOTAL** | | | | | **70.18** |

Table 1. Size-resolved distribution of irradiation-induced dislocation loops within the simulated volume (20 × 20 × 20 µm³). The number of layers, total loop count, and corresponding number of loops intersecting a representative slip plane (per 400 µm²) and calculated loop-induced hardening stress $\tau_i$ are reported for each loop diameter class. The number of loops per slip plane $L_i$ is normalized by the respresentative area to obtain the corresponding areal density (loops/µm²).

#### 4.2.3.2 Contribution from Re-decorated voids

In addition to loops, nanoscale Re/Os-decorated voids (5–10 nm) with a number density *N* of 5 x 10 $^{21}$ m$^{-3}$ , in the size range 5–10 nm, was also reported in the defect microstructure of the irradiated material (Lloyd et al., 2022). This yields $N_v$ of approximately $4 \times 10^7$ voids. The TEM data analysis reports void diameters to lie. Representative diameters of voids considered were 5, 6, 7, 8, and 9 nm, producing the size-resolved counts $N_i$ $L_i$, and $\rho_i$ $(here\ stated\ as\ \boldsymbol{\rho}_{\mathbf{exp},i})$ listed in Table 2.

Osetsky et al. has used MD simulations to compute the pinning stresses associated with voids of varying sizes in tungsten (Osetsky, 2021). However, these are voids without any rhenium or osmium decoration. In the absence of MD data for decorated voids, the values of (Osetsky, 2021) were adopted; Re decoration has been shown to have only a modest effect on the interaction of edge dislocations with Re-decorated dislocation loops (Yang et al., 2023), and a similar insensitivity is assumed here for voids. Molecular dynamics simulations provide the corresponding unpinning stresses $\tau_i$ for each void size: approximately 500 MPa (5 nm), 550 MPa (6 nm), 580 MPa (7 nm), 625 MPa (8 nm), and 660 MPa (9 nm) (9 nm is the largest size for which MD unpinning stresses are available).

The MD unpinning stresses correspond to the obstacle spacing imposed by the periodic simulation cell, which differs from the obstacle spacing in the irradiated material. To transfer them to the experimental microstructure, they were rescaled using dispersed barrier hardening theory, in which, for a given obstacle strength, the obstacle-induced stress scales with the inverse of the obstacle spacing, and hence with the square root of the areal obstacle density:

$$\boxed{\tau_i^{\text{adjusted}} = \tau_i^{\text{MD}} \sqrt{\frac{\rho_i^{\text{cut}}}{\rho_{\text{MD},i}}}} \quad (26)$$

where $\tau_{\text{MD},i}$ is the atomistic unpinning stress, $\rho_{\text{exp},i}$ is the experimentally determined cutting density for that size class, and $\rho_{\text{MD},i}$ is the areal obstacle density of the corresponding MD cell. (Osetsky, 2021) varied the cell dimensions with void size to limit boundary effects, so that $\rho_{\text{MD}}$ decreases from $5.6 \times 10^{14}$ m$^{-2}$ for the smallest void simulated (1 nm) to $1.7 \times 10^{14}$ m$^{-2}$ for the 9 nm void. Values of $\rho_{\text{MD},i}$ for the five size classes were obtained by linear interpolation between these limits (Table 2). The resulting density-adjusted stresses are listed in Table 2.

Because the five size classes represent distinct populations of penetrable obstacles acting concurrently on the glide plane, their contributions were combined using root-sum-square superposition:

$$\tau_{\text{void}} = \sqrt{\sum_{i=1}^{5} \tau_{\text{adjusted},i}^2} \quad (27)$$

giving $\tau_{\text{voids}} \approx 222.2$ MPa

| Size (nm) | $N_i$ | Layers | $L_i$(per 400 μm²) | $\rho_{\exp,i}$ ($m^{-2}$) | MD density, $\rho_{MD,i}$ ($m^{-2}$) | $\tau_{MD,i}$ (MPa) | Adjusted stress, $\tau_{adjusted}$ (MPa) |
|---|---|---|---|---|---|---|---|
| 5 | $2.18 \times 10^6$ | 4000 | 545 | $1.36 \times 10^{12}$ | $3.650 \times 10^{14}$ | 500 | 30.52 |
| 6 | $9.77 \times 10^6$ | 3333 | 2930 | $7.33 \times 10^{12}$ | $3.163 \times 10^{14}$ | 550 | 83.73 |
| 7 | $1.61 \times 10^7$ | 2857 | 5636 | $1.409 \times 10^{13}$ | $2.675 \times 10^{14}$ | 580 | 133.09 |
| 8 | $9.77 \times 10^6$ | 2500 | 3908 | $9.77 \times 10^{12}$ | $2.188 \times 10^{14}$ | 625 | 132.08 |
| 9 | $2.18 \times 10^6$ | 2222 | 981 | $2.45 \times 10^{12}$ | $1.700 \times 10^{14}$ | 660 | 79.23 |
| RSS Total | – | – | | | | | 222.2 |

Table 2. Size-resolved distribution of Re/Os-decorated nanovoids used for constitutive model calibration. For each void size class, the total number of voids ($N_i$), corresponding number of layers, projected number of voids intersecting a slip plane ($L_i$ per 400 μm²), experimental planar obstacle density $\rho_{\exp,i}$, atomistically determined critical resolved shear stress $\tau_{MD,i}$ and $\tau_{adjusted}$ density-adjusted stress contribution are reported

#### 4.2.3.3 Contribution from Re and Os precipitates

The third defect population consists of Re/Os-rich precipitates, with $N$ of $2.5 \times 10^{22}$ $m^{-3}$ and diameters of 4–10 nm (Lloyd et al., 2022). This gave $N_v$ of $2.0 \times 10^8$.

Since MD unpinning stresses are available for coherent Re-rich precipitates of 4.4, 6.6, 7.7, 8.8 and 9.0 nm diameter (Osetsky, 2022). These five sizes were therefore adopted as representative classes. $N_i$ was obtained using a normal distribution (mean 7.3 nm, standard deviation 1 nm). $N_i$ and associated $L_i$, and $\rho_i$ are stated in Table 3.

| Size | $N_i$ | Layers | $L_i$ (per plane) | Experimental density, $\rho_i$ ($m^{-2}$) | MD CRSS, $\tau_{MD}$ (MPa) | Adjusted stress, $\tau_{adjusted}$ (MPa) |
|---|---|---|---|---|---|---|
| 4.4 | $1.06\times10^{6}$ | 4545 | 233 | $5.82\times10^{11}$ | 501 | 16.05 |
| 6.6 | $6.85\times10^{7}$ | 3030 | $2.26\times10^{4}$ | $5.65\times10^{13}$ | 610 | 192.58 |
| 7.7 | $8.08\times10^{7}$ | 2597 | $3.11\times10^{4}$ | $7.78\times10^{13}$ | 641 | 237.46 |
| 8.8 | $2.84\times10^{7}$ | 2273 | $1.25\times10^{4}$ | $3.13\times10^{13}$ | 644 | 151.32 |
| 9.0 | $2.12\times10^{7}$ | 2222 | $9.54\times10^{3}$ | $2.39\times10^{13}$ | 692 | 142.09 |
| RSS TOTAL | | | | | | 370 |

Table 3. Size-resolved quantification of strengthening contributions from irradiation-induced precipitates. The total population $N_i$, corresponding number of layers, and projected number of precipitates intersecting a representative slip plane $L_i$ experimental planar obstacle density ($\rho_{\mathrm{exp},i}$), atomistically determined critical resolved shear stress ($\tau_{\mathrm{MD},i}$), and density-adjusted stress contribution ($\tau_{\mathrm{adjusted},i}$) are reported.

The MD unpinning stresses are 501, 610, 641, 644 and 692 MPa for the 4.4, 6.6, 7.7, 8.8 and 9.0 nm precipitates, respectively (Osetsky, 2022). No corresponding data exist for Os-rich precipitates. As Re and Os are adjacent in the periodic table and form structurally and chemically similar clusters in irradiated tungsten, the Re-derived strengths were applied to all precipitates. In the MD simulations, a fixed dislocation segment length of L = 42 nm was used for all sizes, so that the periodic cell corresponds to a constant areal obstacle density of

$$\rho_{\mathrm{MD}} = \frac{1}{L^2} \approx 5.669 \times 10^{14}\ \mathrm{m}^{-2}$$

The MD stresses were scaled to the experimental densities using Eq. (26), and the size-resolved contributions (Table 3) were combined by root-sum-square superposition (Eq. 32).

$$\tau_{\text{ppt}} = \sqrt{\sum_{i=1}^{5} \tau^2_{\text{adjusted},i}} \quad (32)$$

Therefore, the total effective precipitate-induced hardening contribution, on the active glide plane is:

$$\boxed{\tau_{\text{ppt}} \approx 370\ \text{MPa}}$$

The MD-derived critical stresses in Sections 4.2.3.2 and 4.2.3.3 are used as atomistic estimates of obstacle strength rather than as directly transferable continuum parameters. The MD simulations were performed at strain rates of approximately $10^6$ $s^{-1}$ (Osetsky, 2022, 2021), far higher than those of the present experiments and CPFE simulations. However, prior MD studies show that the calculated obstacle strength varies little at the lower end of the MD strain-rate range, below approximately $10^7$ $s^{-1}$ (Das et al., 2024; Mishra and Das, 2026), supporting their use as rate-independent obstacle strengths, consistent with the athermal treatment of $\tau_H$ in the constitutive model. The density scaling of Eq. (26) accounts for the obstacle spacing imposed by the MD cell, but residual finite-cell effects, such as line-tension and image interactions, cannot be fully removed, so the resulting values represent idealised estimates of obstacle strength.

#### 4.2.3.4 Effective macroscopic resistance and superposition of defect contributions

So far we have obtained the strengthening contributions from the irradiation-induced defect populations namely, dislocation loops, nanovoids, and Re/Os-rich precipitates.—The superposition of strengthening contributions from different obstacle populations is not unique. Here we first adopted a RSS combination of the three contributions i.e.

$$\tau_H^0 = \sqrt{\tau_{loop}^2 + \tau_{void}^2 + \tau_{ppt}^2}$$

This yielded ≈ 437 MPa. However, with this $\tau_H^0$, $\gamma_s$ could not be tuned to reproduce the experimentally measured load–displacement and pile-up response of the irradiated material at the calibration strain rate. Thus, a linear sum i.e.

$$\tau_H^0 = \tau_{\text{loop}} + \tau_{\text{void}} + \tau_{\text{ppt}} \quad (33)$$

was adopted. This represents a simplifying assumption rather than a unique physical description of multi-obstacle strengthening, and the resulting $\tau_H^0$ should be regarded as an effective obstacle resistance constrained jointly by the microstructural estimate and by the indentation response.

This yields an effective strengthening of approximately:

$$\tau_H^0 \approx 662 \text{ MPa}.$$

The loop contribution depends on the assumed interaction geometry; the limiting values m = 1 and m = 2 give $\tau_H^0 \approx 639$ and 686 MPa, respectively, and simulations with $639 < \tau_H^0 < 686$ MPa reproduced the experimental response equally well, so the results are insensitive to this choice. We note here that in the above-stated calculations, the defect populations are treated as statistically homogeneous, and the hardening is defined through obstacle densities, so the quarter-symmetry finite-element model does not alter the prescribed defect distribution. The populations are evaluated independently and combined in a mean-field sense. Thus, spatial correlations between defects and simultaneous interactions of a dislocation with several defect types are not resolved.

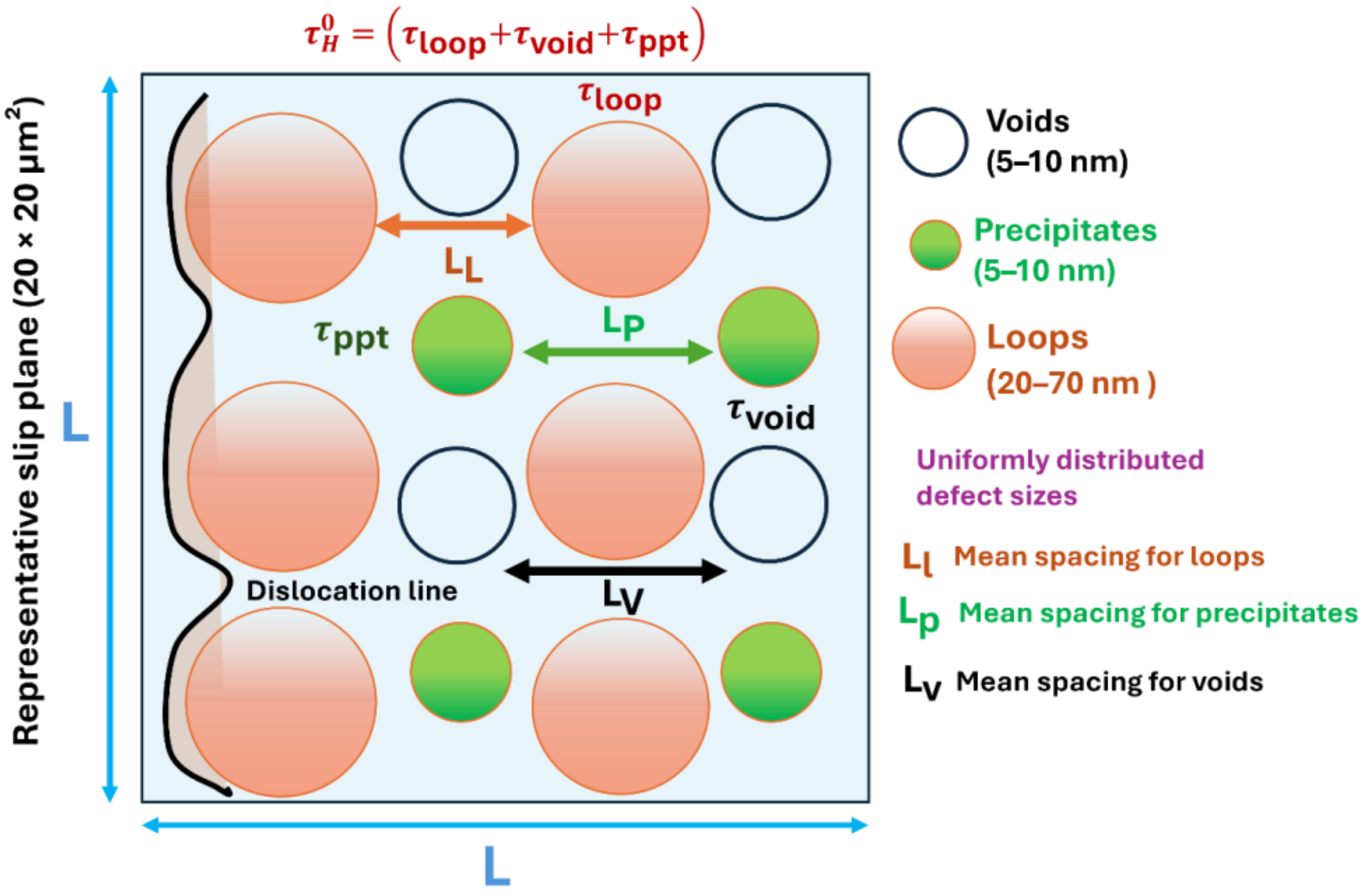

**Fig. 3**. Schematic of the estimate of the irradiation-induced resistance $\tau_H^0$ (not to scale). Dislocation loops (20–70 nm), Re/Os-decorated voids (5–10 nm) and Re/Os-rich precipitates (4–10 nm) are distributed uniformly in space within the 20 × 20 × 20 µm³ representative volume and projected onto a representative 20 × 20 µm² slip plane; one characteristic size is shown for each population, although the sizes are allocated to discrete classes using a normal distribution. $L_L$, $L_V$ and $L_P$ denote the mean obstacle spacings of the loops, voids and precipitates, which determine the stresses $\tau_{loop}$, $\tau_{void}$ and $\tau_{ppt}$ required for a gliding dislocation to overcome each population (Sections 4.2.3.1–4.2.3.3). These are combined according to Eq. (33) to give $\tau_H^0$ (Section 4.2.3.4). In the present mean-field treatment each population is evaluated independently, although a dislocation may interact with several obstacles of different types simultaneously.

### 4.3 Model Calibration Summary

The model was calibrated against the experimental response at the highest nominal strain rate ($3.2 \times 10^{-3}$ s$^{-1}$). For the unirradiated material, two parameters, $\tau_c^0$ and $C'$, were fitted to the load–displacement response (to the peak load) and pile-up. For the irradiated material, these parameters were retained unchanged, and irradiation effects were introduced through two additional parameters: the initial irradiation-induced resistance $\tau_H^0$, obtained from the defect microstructure (Section 4.2.3), and the softening parameter $\gamma_s$, fitted to the irradiated response at the same strain rate. Because each indent terminated at the depth at which the target indentation strain was reached, peak loads are not directly comparable between indents.

Calibration was therefore performed against a representative indent for each condition. (The indent-to-indent repeatability was characterised by through yield stresses extracted from the indentation stress–strain curves of 4–9 successful indents per condition: the standard deviation was approximately 1–4% of the mean for unirradiated tungsten and approximately 4–10% for irradiated tungsten (Jim, n.d.). All parameters were then held fixed, and the simulations at the two lower strain rates were performed to the corresponding experimentally achieved depths, changing only the prescribed displacement rate. These simulations were assessed against the measured load–displacement response, pile-up, residual lattice strain and GND distributions.

## 5. Results

### 5.1 Load-displacement curves

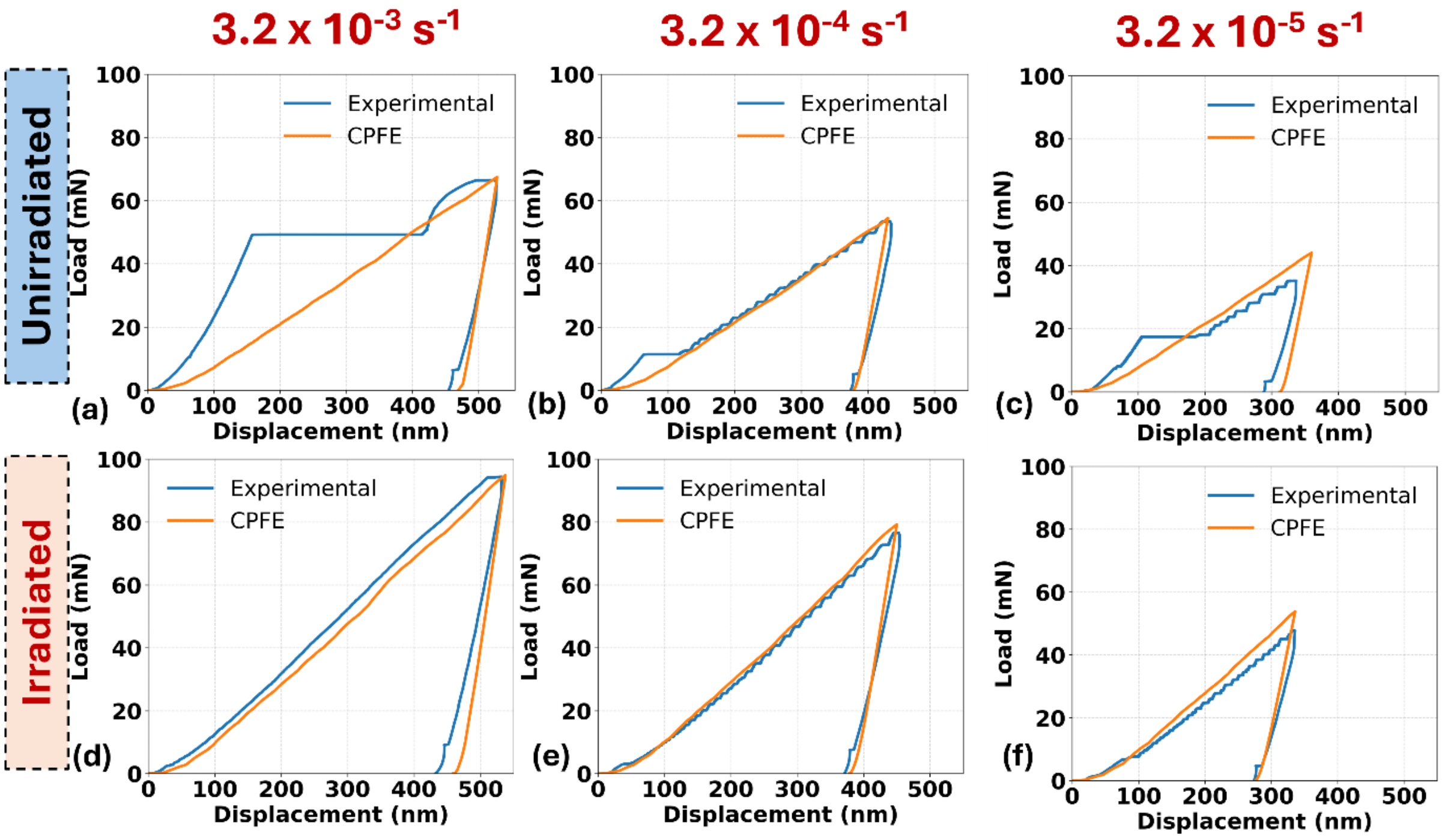

**Fig. 4**. Load-displacement curves as measured by nano-indentation and predicted by CPFE for unirradiated and neutron-irradiated single-crystal tungsten for strain rates 3.2 × $10^{-3}$ $s^{-1}$, 3.2 × $10^{-4}$ $s^{-1}$, and 3.2 × $10^{-5}$ $s^{-1}$.

**Fig. 4** compares the experimentally measured and CPFE-predicted load-displacement responses for unirradiated and neutron-irradiated tungsten across the three investigated strain-rate conditions. As discussed in Section 3, the strain-rate control protocol resulted in different maximum indentation depths being attained at each nominal strain rate. These experimentally achieved depths were reproduced exactly in the corresponding CPFE simulations to ensure direct comparison between model predictions and measurements.

For both material conditions, the maximum indentation load decreases systematically from the fastest to the slowest nominal strain-rate condition, together with the final depth reached. Because the depth differs between tests, this reduction cannot be attributed to intrinsic rate sensitivity alone; the contributions of strain rate and depth are separated in Section 5.4. At a given nominal strain-rate condition, the irradiated material consistently sustains substantially higher loads than the unirradiated material. This reflects the increased resistance to plastic deformation arising from irradiation-induced defect populations.

Discrete displacement bursts are observed in the unirradiated material, particularly at the higher strain-rate conditions. These events are characteristic of intermittent dislocation activity in initially defect-sparse crystals and are commonly associated with the activation of discrete dislocation sources during the early stages of plastic deformation (Biener et al., 2007). Pronounced pop-ins are largely absent in the irradiated material, where the dense population of dislocation loops, Re/Os-rich precipitates and decorated voids may be providing abundant sites for dislocation nucleation, suppressing large displacement excursions (Das et al., 2020). As such stochastic dislocation events are not represented within the CPFE framework, these pop-ins are not reproduced in the simulated responses.

Apart from the pop-ins, the CPFE simulations reproduce the load–displacement response for all conditions, capturing the elastic loading, the slope of the loading curve and the maximum load at the experimentally achieved depth (with the largest difference, about 25%, for unirradiated tungsten at the slowest strain rate). The constitutive parameters were calibrated at the highest strain rate only and applied unchanged at the two lower strain rates. The calibrated framework therefore reproduces the indentation response at the experimentally achieved depths across all three nominal strain rates without further parameter adjustment.

## 5.2 Surface morphologies from AFM

**Fig. 5** compares the AFM surface morphologies with the CPFE predictions for both materials at the three nominal strain rates, and Fig. A.1 compares line profiles extracted along ⟨110⟩, each the average of two diametrically opposite lobes within ±5° of that direction. In all cases, deformation produces the characteristic four-fold pile-up pattern around the indent, with lobes along ⟨110⟩ and little material displaced along ⟨100⟩, reflecting the crystallographic anisotropy of slip in the (001)-oriented crystal.

The CPFE simulations reproduce this morphology in all six conditions, including the four-fold symmetry, the alignment of the lobes along ⟨110⟩ and their radial position relative to the indent. The line profiles (Fig. A.1) show good quantitative agreement between the simulation predictions and the experimental observations. For unirradiated tungsten, the predicted pile-up heights agree with the measurements to within about 10 nm at all three strain rates. For irradiated tungsten, the measured lobes are asymmetric at the two faster strain rates, whereas the simulated profiles are symmetric; the predicted pile-up heights lie between the two measured lobes. This asymmetry, which likely reflects local microstructural heterogeneity, is not encoded in the model. This is why the profiles of two opposite lobes were averaged for model calibration (**Appendix A**).

Irradiation markedly increases pile-up. At each nominal strain rate, the pile-up height in irradiated tungsten is about 1.5–2.3 times that in unirradiated tungsten, with steeper flanks around the indent, and the model captures this increase. This is consistent with the confinement of plastic flow by irradiation-induced defects reported previously for ion-implanted tungsten (Das et al., 2020, 2019a) obstacles restrict dislocation motion and the lateral expansion of the plastic zone, so a greater fraction of the displaced material is accommodated at the free surface. In both materials, pile-up also decreases from the fastest to the slowest nominal strain rate. Relative to the slowest condition, the fastest-rate pile-up is about 66% higher in unirradiated and about 150% higher in irradiated tungsten. As the final depth also decreases between these tests, these differences cannot be attributed to strain rate alone. The contributions of depth and rate are separated in Section 5.4.

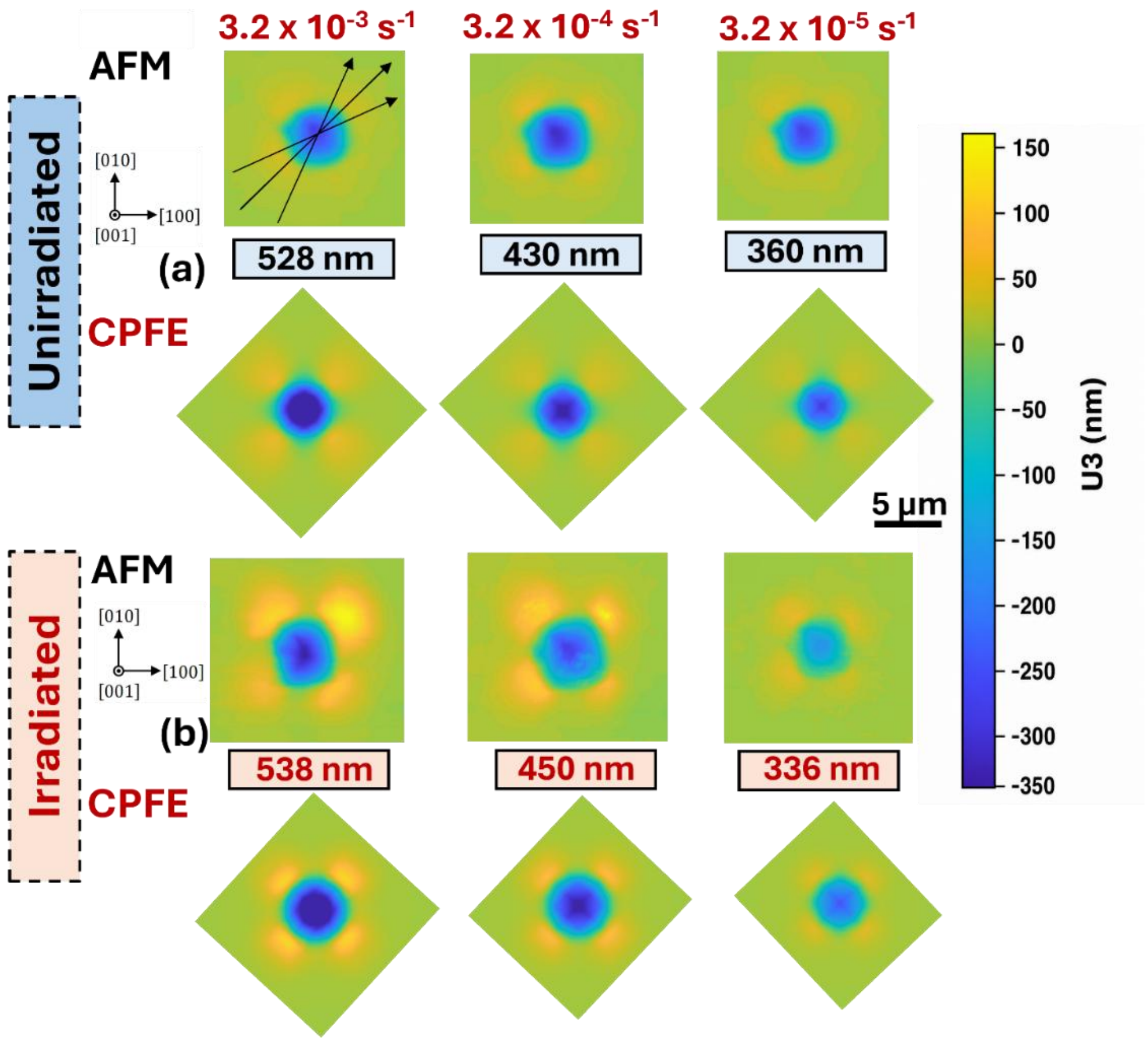

**Fig. 5.** Surface morphologies (height profile) of nano-indents in (a) pure tungsten and (b) neutron-irradiated tungsten as predicted by CPFE and measured by AFM after indentation, labeled with maximum indentation depth reached for strain rates 3.2 x $10^{-3}$ $s^{-1}$, 3.2 x $10^{-4}$ s-1, and 3.2 x $10^{-5}$ $s^{-1}$. The black arrows in (a) indicate the directions along which AFM line profiles were extracted; profiles from two diametrically opposite lobes within ±5° of the ⟨110⟩ direction were averaged for quantitative comparison and CPFE calibration.

### 5.3 Elastic strain fields and GND density maps

To examine the deformation fields beyond the surface, the residual elastic strains and GND densities around the indents were measured by HR-EBSD and compared with the CPFE predictions. As HR-EBSD resolves the deviatoric elastic strain (Section 2.5), the deviatoric part of the CPFE elastic strain ($\varepsilon^e_{dev}$) was extracted for comparison:

$$\varepsilon^e_{dev} = \varepsilon^e - \varepsilon^e_{vol} = \varepsilon^e - 1/3\ \mathrm{Tr}\,(\varepsilon^e)\,\boldsymbol{I} \quad (34)$$

where $\varepsilon_{vol}^{e}$ is the volumetric strain. The CPFE strain and GND fields were extracted on the free top surface (XY plane) of the three-dimensional model after unloading, corresponding to the near-surface layer sampled by HR-EBSD.

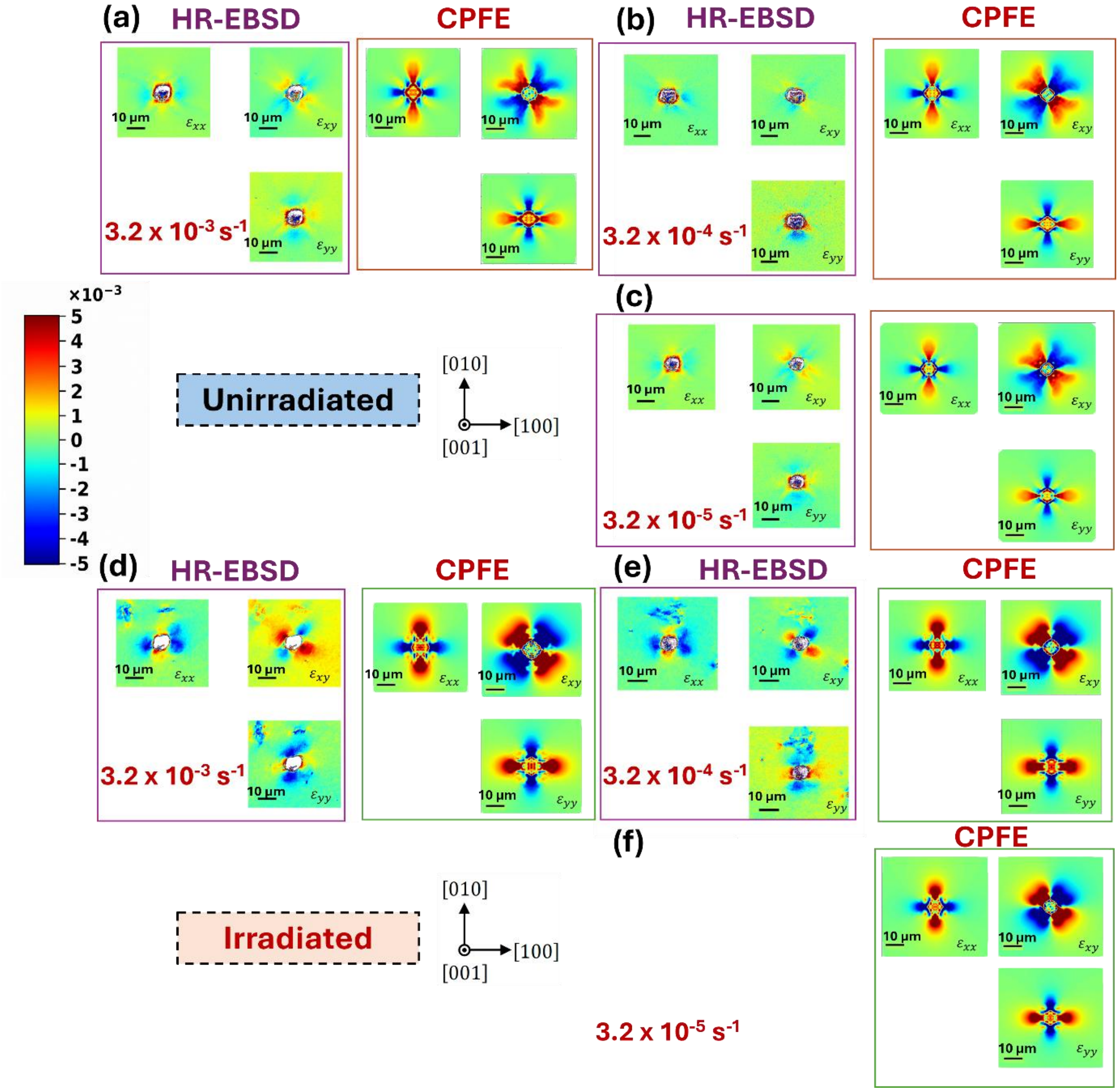

**Fig. 6.** Residual elastic deviatoric lattice strain plotted on the XY plane (sample surface) as predicted by CPFE and measured by HR-EBSD after indentation for strain rates 3.2 x $10^{-3}$ $s^{-1}$, 3.2 x $10^{-4}$ $s^{-1}$, and 3.2 x $10^{-5}$ $s^{-1}$. The HR-EBSD data at the lowest strain rate for irradiated tungsten was partially damaged and is therefore not shown in the figure.

**Fig. 6** compares the measured and predicted residual strain maps, and Fig. B.1 compares line profiles taken horizontally through the indent centre. In both materials, $\varepsilon_{xx}$ and $\varepsilon_{yy}$ show the characteristic pattern of radial compression and circumferential tension around the indent, and $\varepsilon_{xy}$ shows a four-lobed distribution consistent with crystallographically anisotropic deformation of the (001) crystal. The CPFE reproduces the sign, symmetry and orientation of all three strain components in both materials. Agreement is closest for irradiated tungsten, where the model captures the position and shape of the compressive $\varepsilon_{xx}$trough immediately outside the indent, although it somewhat overestimates its magnitude. For unirradiated tungsten, the model overestimates the residual strain outside the indent more substantially, by up to a factor of about five, and in both materials it predicts broader tensile $\varepsilon_{yy}$ lobes than measured (Appendix B). Strains within the indent cavity, where pattern quality is poor, are not compared.

Irradiation increases and localises the residual elastic strain around the indent. In irradiated tungsten, the measured compressive strain is several times larger than in unirradiated tungsten and peaks immediately outside the indent edge, whereas in unirradiated tungsten it is weaker and peaks further out. Although the strain decays over a similar distance in both materials, it falls off more steeply in the irradiated material, so the residual strain is concentrated closer to the indent. This is consistent with irradiation-induced voids, dislocation loops and Re/Os-rich precipitates restricting dislocation motion and confining the plastic zone, and agrees with the larger pile-up in the irradiated material (Section 5.2).

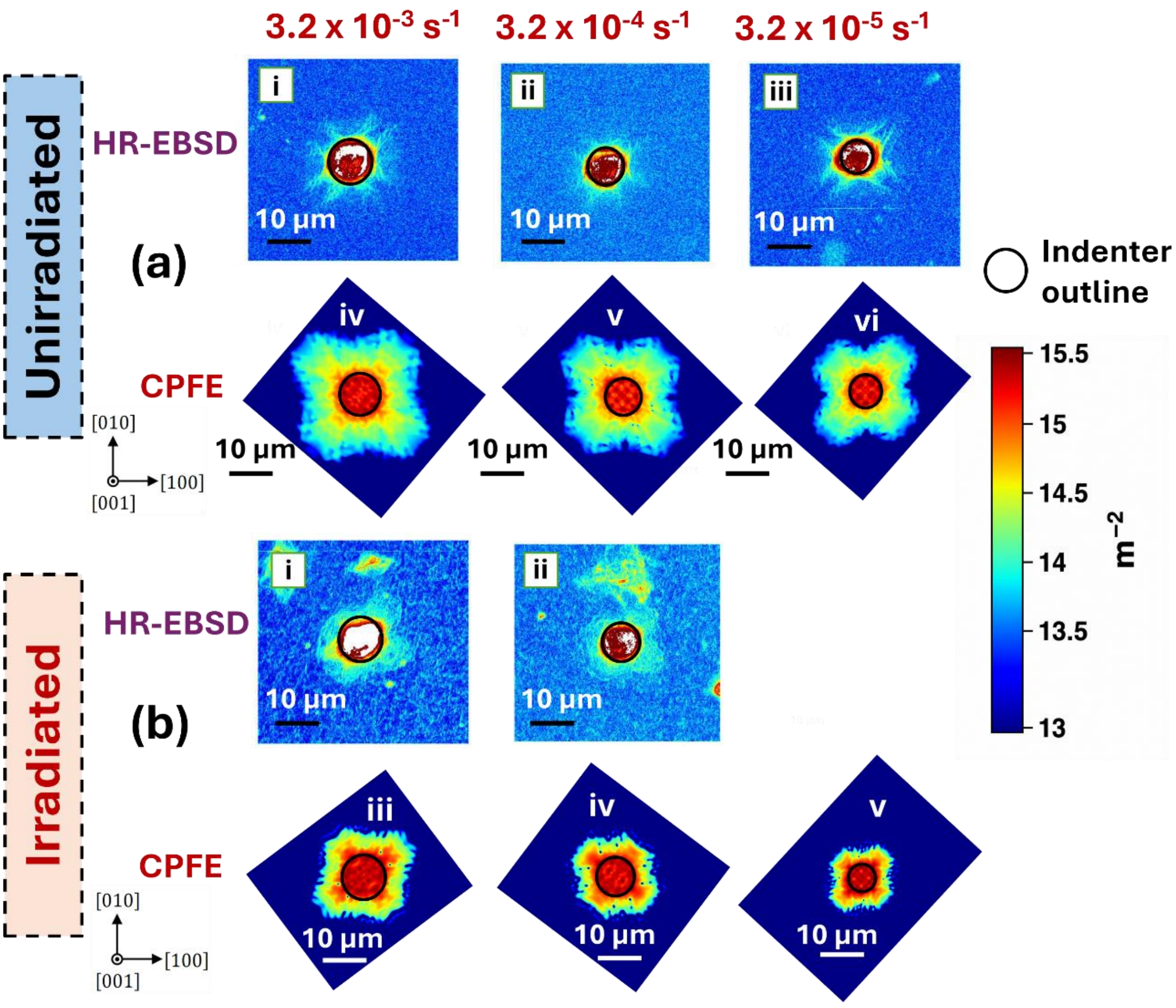

**Fig. 7.** Sum of GNDs (ρ) over all 24 slip systems as predicted by CPFE and measured by HR-EBSD for (a) pure tungsten and (b) neutron-irradiated tungsten for strain rates 3.2 x $10^{-3}$ $s^{-1}$, 3.2 x $10^{-4}$ $s^{-1}$, and 3.2 x $10^{-5}$ $s^{-1}$. For unirradiated tungsten, panels (i), (ii), and (iii) show the ρ via HR-EBSD measurements at 3.2 x $10^{-3}$ $s^{-1}$, 3.2 x $10^{-4}$ $s^{-1}$, and 3.2 x $10^{-5}$ $s^{-1}$, respectively, while panels (iv), (v), and (vi) show the corresponding CPFE predictions. For neutron-irradiated tungsten, panels (i) and (ii) show the ρ via HR-EBSD measurements at 3.2 x $10^{-3}$ $s^{-1}$ and 3.2 x $10^{-4}$ $s^{-1}$, respectively, while panels (iii), (iv), and (v) show the corresponding CPFE predictions. The plots are shown on XY plane (sample surface). The same colour scale is used for all plots showing log10(ρ), with ρ in 1/$m^2$.

**Fig 7.** compares the measured and predicted GND densities (Fleck and Hutchinson, 1997; Nye, 1953) for unirradiated and neutron-irradiated tungsten across the investigated strain-rate range. We note here that the HR-EBSD data at the lowest strain rate was partially damaged and is therefore not shown in the figure.

Details of the GND density computation are provided elsewhere (Suchandrima Das et al., 2018). In this work, the L2 minimization approach is employed, wherein the sum of the squares of the individual dislocation densities is minimized to determine the total GND density. This calculation is performed by summing the contributions from the 12 $a/2\langle 111\rangle\ \{110\}$ and 12 $a/2\langle 111\rangle\ \{112\}$ slip systems considered in this study (Marichal et al., 2013; Srivastava et al., 2013). The total GND density from CPFE simulations is then compared with experimentally measured values from high-resolution EBSD (HR-EBSD).

In both materials, the measured GND density is highest at the indent edge and decreases outwards, with a four-fold symmetry about the indent. The simulations reproduce the four-fold symmetry and the decrease of GND density away from the indent seen in the measured fields. Following neutron irradiation, there is an increase in the GND density field magnitude, but it remains confined closer to the vicinity of the indent. This behaviour is consistent with the strain and pile-up measurements presented previously, which indicated increasingly localised deformation following irradiation. The simulated GND field is likewise more compact after irradiation, consistent with this localisation.
Smaller differences are also observed between the nominal strain-rate conditions. For both material conditions, the GND distributions become progressively more localised from the fastest to the slowest condition, with the fastest condition exhibiting the most spatially distributed deformation field.

We note here that the absolute magnitudes of GND densities derived from HR-EBSD and CPFE differ, with CPFE predictions exceeding the experimentally measured values in certain areas of the GND contour map (fig. 7). Such discrepancies may be occurring because GND density determination is inherently dependent on spatial resolution and the method used to evaluate lattice curvature (Ashby, 1970). In addition, HR-EBSD measurements are affected by pattern noise, cross-correlation uncertainty and spatial averaging within the electron interaction volume, all of which tend to smooth steep strain and rotation gradients in highly deformed regions. Consequently, the experimentally derived lattice curvature, and hence the inferred GND density, may be underestimated relative to the numerically resolved fields. Furthermore, only a subset of the full Nye tensor is experimentally accessible from HR-EBSD measurements (Pantleon, 2008), whereas the CPFE framework provides complete access to the lattice distortion field. Quantitative differences in GND magnitude are therefore expected. Nevertheless, the agreement in the morphology and symmetry of the GND fields, and in their variation between the tested conditions, indicates that the CPFE framework captures the spatial organisation of the deformation beneath the indent.

## 5.4 CPFE guided experiments

Taken together, the CPFE framework reproduces the load–displacement response and pile-up at all three nominal strain rates, and the principal features of the residual strain and GND fields. Importantly, however, the experimental observations alone cannot uniquely isolate the intrinsic effect of strain rate. As discussed previously, the strain-rate control protocol resulted in different maximum indentation depths being attained at different loading rates, such that the measured variations in load response, pile-up morphology, lattice strain and GND density reflect the combined influence of strain rate, indentation depth and evolving deformation geometry. Having been calibrated and assessed against these independent experimental observables, the CPFE framework provides a controlled platform for separating these coupled effects. In the following section, the model is therefore used to vary strain rate and indentation depth independently, within the adopted constitutive framework, to interpret the experimentally observed response.

Here, the assessed CPFE framework is employed to conduct controlled simulations designed to decouple the effects of strain rate and indentation depth. First, simulations are performed at a fixed indentation depth of 500 nm while varying the applied strain rate, enabling direct assessment of intrinsic strain-rate sensitivity under depth-controlled conditions. Subsequently, complementary simulations are carried out at a constant strain rate while varying the prescribed indentation depth, allowing the independent influence of deformation geometry to be examined. Together, these model-based studies provide a systematic interpretation of the coupled strain-rate and depth effects that could not be isolated experimentally.

### 5.4.1 Deconvolution of the effect of strain rate on peak load and peak pile-up

**Fig. 8** presents the CPFE-predicted variation of peak load and peak pile-up with strain rate for unirradiated and neutron-irradiated tungsten. In contrast to the experiments, all simulations were performed to a fixed indentation depth of 500 nm, removing the influence of depth-dependent geometric effects so that the contribution of strain rate can be examined independently. The strain-rate range considered, $3.2 \times 10^{-5}$ to $3.2 \times 10^{1}$ $s^{-1}$, extends beyond that accessible experimentally. However, the constitutive formulation is based on thermally activated dislocation-obstacle interactions, for which the governing activation energy is assumed to remain unchanged across the investigated strain-rate range. Previous mechanistic studies have shown that such formulations remain applicable over several orders of magnitude in strain rate, with the rate sensitivity arising from the competition between loading time and thermally activated obstacle escape rather than from changes in the activation energy (Zheng et al., 2016).

The peak load response (**Fig. 8a**) shows that irradiation increases the resistance to deformation across the entire strain-rate range, consistent with the presence of irradiation-induced obstacles. For both materials, the peak load is essentially insensitive to strain rate between $3.2 \times 10^{-5}$ and $3.2 \times 10^{-1}$ $s^{-1}$, varying by only a few percent. A clear increase is predicted only at the highest strain rate, where the reduced time available for thermal activation raises the flow resistance (Dezerald et al., 2015; Fitzgerald, 2016). This increase is larger for unirradiated tungsten (about 30–40%) than for irradiated tungsten (about 9%). This weaker rate dependence of the irradiated material follows from the athermal treatment of the irradiation-induced resistance in the model, and is discussed in Section 6.3.

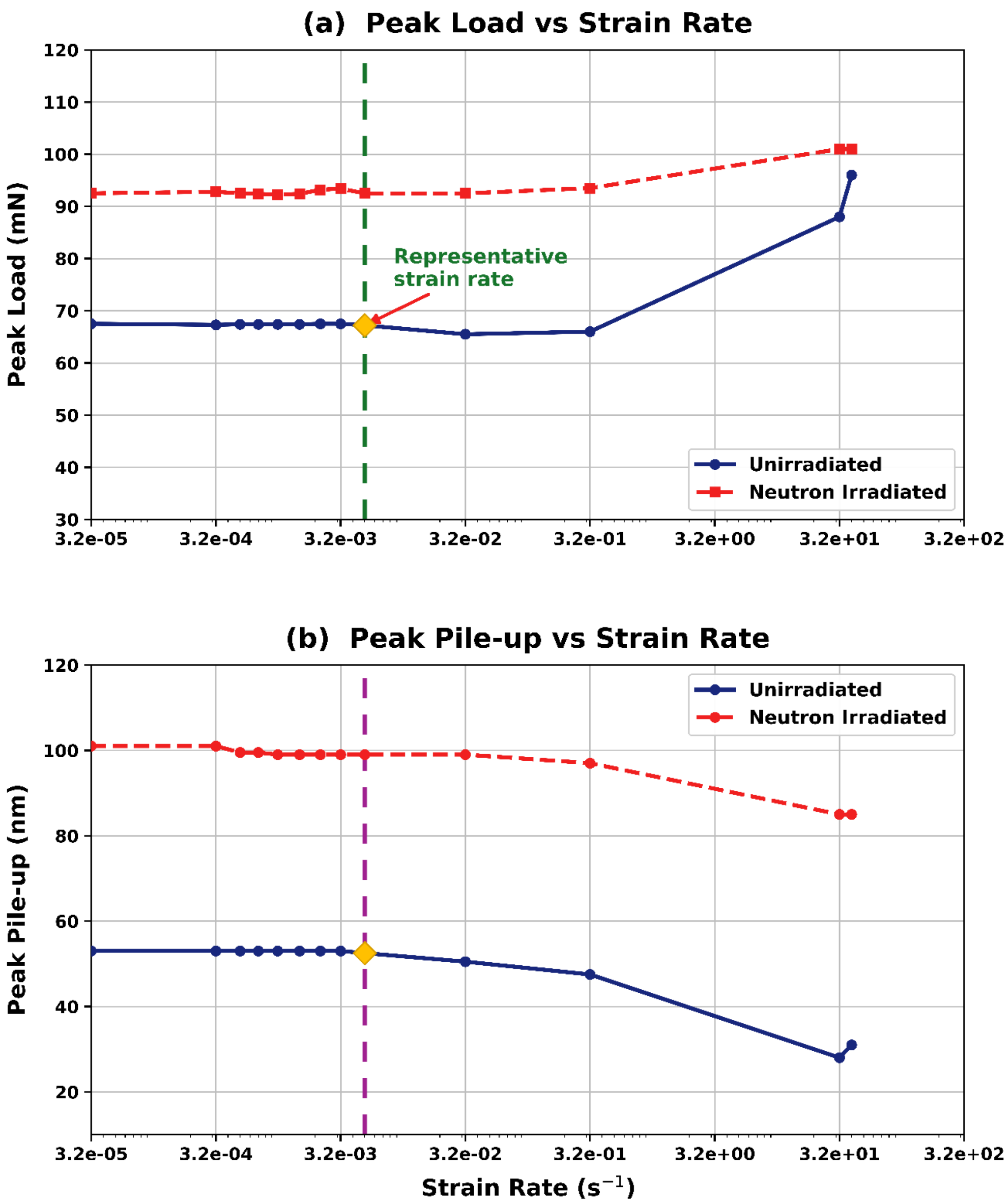

**Fig. 8**. Effect of strain rate at a fixed depth of 500nm on the plastic response of tungsten under nanoindentation. (a) Variation of peak load with strain rate for unirradiated and neutron-irradiated tungsten. (b) Variation of peak pile-up with strain rate under identical conditions.

The pile-up response (**Fig. 8b**) shows the opposite trend: in both materials, peak pile-up decreases with increasing strain rate. This is consistent with the higher flow stress at faster loading, which reduces the plastic flow of material towards the free surface. At all strain rates, irradiated tungsten shows substantially larger pile-up than unirradiated tungsten, consistent with the more localised deformation observed experimentally (Sections 5.2 and 5.3). As with the load response, the variation in pile-up is small over the low and intermediate strain-rate range and becomes significant only at the highest rates.

These fixed-depth simulations indicate that, within the assumptions of the adopted constitutive framework, the intrinsic effect of strain rate is small within the experimental strain-rate range. Between $3.2 \times 10^{-5}$ and $3.2 \times 10^{-3}$ $s^{-1}$, the predicted peak load changes by less than 1% and the peak pile-up by less than 2 nm in both materials. A representative strain rate can therefore be selected for the simulations investigating indentation-depth effects (Section 5.4.2), allowing the influence of deformation length scale to be examined independently of strain rate.

### 5.4.2 Deconvolution of the effect of Indentation depth on peak load and peak pile-up (at constant strain rate of 5 x$10^{-3}$ $s^{-1}$)

**Fig. 9** presents the CPFE predictions of peak load (**Fig. 9a**) and peak pile-up (**Fig. 9b**) as functions of prescribed indentation depth, from 100 to 1400 nm, for unirradiated and neutron-irradiated tungsten. All simulations were performed at a constant strain rate of $5 \times 10^{-3}$ $s^{-1}$. This lies within the rate-insensitive regime identified in Section 5.4.1, where the response differs by less than 1% from that at $3.2 \times 10^{-3}$ $s^{-1}$, so the influence of deformation length scale is isolated. Indentation depth was varied because analytical models and experimental observations have shown that increasing penetration progressively expands the plastic zone beneath the indenter, modifying the extent of dislocation activity and strain gradients (Chen and Bull, 2006; Feng and Nix, 2004; Hardie et al., 2015). In both materials, the peak load and the peak pile-up are seen to increase monotonically with depth, as deeper indentation activates a larger plastically deformed volume, requiring higher loads and displacing more material towards the surface.

Irradiation increases both quantities at every depth, but the two respond differently. The peak load of irradiated tungsten is about 30–50% higher than that of unirradiated tungsten across the entire depth range, from about 10 versus 7 mN at 100 nm to about 246 versus 190 mN at 1400 nm. The relative irradiation effect on load is therefore approximately independent of depth, although the absolute difference grows with the load. The irradiation effect on pile-up, in contrast, is itself depth dependent. The ratio of irradiated to unirradiated pile-up increases from about 1.6 at 200 nm to about 2 at 500–700 nm, and remains approximately constant at greater depths, reaching about 370 versus 195 nm at 1400 nm. The irradiation-induced difference in pile-up therefore becomes more pronounced as the depth increases towards about 500 nm.

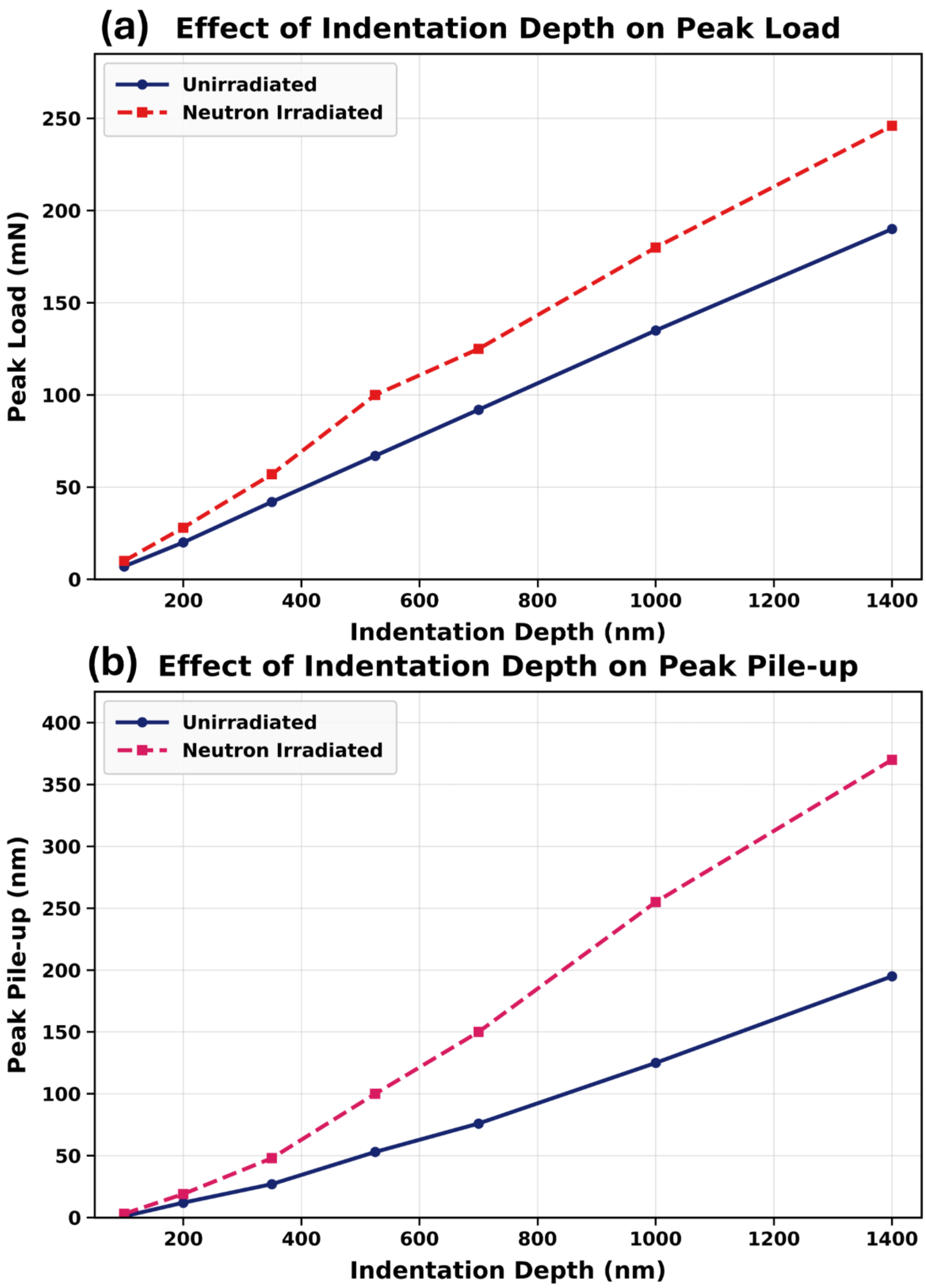

**Fig. 9.** Effect of indentation depth at a fixed strain rate of $5 \times 10^{-3}$ $s^{-1}$ on the nanoindentation response of tungsten. (a) Increase in peak load as a function of indentation depth for unirradiated and neutron-irradiated tungsten. (b) Corresponding variation in peak pile-up with depth.

These fixed-rate simulations therefore show that indentation depth is a critical length scale for the measured response of irradiated tungsten, particularly for the surface deformation. Combined with the fixed-depth simulations, they allow the contributions of loading rate and deformation length scale, which are coupled in the strain-rate-controlled experiments, to be separated. Their implications for the experimental observations are considered next.

### 5.4.3 Synthesis of Results from Virtual Experiments

The results shown above bear directly on the interpretation of the experiments. Within the adopted constitutive framework, the intrinsic rate sensitivity in the experimental range (Fig. 8) is too small to account for the observed differences in final penetration depth (Section 3). Under the strain-feedback protocol, a weakly rate-sensitive material response would produce nearly identical contact responses, and hence similar final depths, at the three nominal strain rates. The depth differences in the present experiments are therefore most likely instrumental or control-related, for example arising from thermal drift during the considerably longer slow tests or from lag in the feedback control, rather than a consequence of material rate sensitivity.

The measured differences between the nominal strain-rate conditions should therefore arise predominantly from the different depths reached, and the fixed-rate simulations (Fig. 9) allow this to be tested. Over the experimental depth ranges (about 360–528 nm for unirradiated and 336–538 nm for irradiated tungsten), depth alone produces increases in pile-up of about 90% and 130%, comparable to the measured increases of about 66% and 150% (Fig. 5). The simulations also reproduce the larger change in irradiated tungsten, which follows from the steeper depth dependence of its pile-up (Section 5.4.2). Depth alone also accounts for most of the measured increase in peak load between the slowest and fastest tests: about 55% and 85% for unirradiated and irradiated tungsten, compared with about 90% and 95% measured (Fig. 4). The remaining difference arises mainly at the slowest strain rate, where the measured peak load lies below the CPFE prediction at the same depth, by about 25% for unirradiated and 12% for irradiated tungsten. This may reflect the instrumental effects noted above, which are most significant in the longest tests. The more compact residual strain and GND fields at the slower conditions (Figs. 6 and 7) are likewise consistent with the smaller contact radii and plastic zones at these depths. Consequently, the agreement between experiment and CPFE at the two lower nominal strain rates primarily reflects the depth-dependent indentation response.

These results can be expressed as a practical criterion for strain-rate-controlled spherical indentation. Over the experimental depth range, the fixed-rate simulations give a peak load that scales approximately as $h^{1.1}$- $h^{1.15}$ and a pile-up that scales approximately as $h^{1.5}$- $h^{1.8}$ , where h is the maximum indentation depth (Fig. 9). A 10% difference in final depth therefore changes the peak load by about 11% and the pile-up by about 15–19%. For the rate sensitivity measured for this material by strain-rate jump tests ($m \approx 0.03$–$0.04$; (Abernethy et al., 2019), the flow stress scales as $\dot{\varepsilon}^m$, so a tenfold increase in strain rate raises it by a factor of $10^m \approx 1.07$–1.10, i.e. by about 7–10%. As the load at a given depth scales approximately with the flow stress, a tenfold change in strain rate changes the load by a similar amount. A 10% difference in final depth thus produces a change in load equivalent to more than an order of magnitude in strain rate, and the depth spread of the present experiments (about 360–528 nm) is equivalent to about five decades, compared with the two decades actually tested. To keep the depth contribution below the effect of a tenfold change in strain rate, final depths would need to agree to within about 7% when comparing loads, and within about 5% when comparing pile-up. These values are specific to the present material and indenter radius, but the approach applies generally: when strain-rate-controlled spherical tests reach different final depths, the depth sensitivity of the measured quantity must be established, for example by depth-controlled simulations as here, before differences between strain-rate conditions can be attributed to rate sensitivity.

# 6 Discussion on Summary of Results

## 6.1 Coupling of strain rate and deformation length scale

Constitutive strain-rate sensitivity and length-scale effects originate from distinct physical mechanisms: thermally activated dislocation motion and strain-gradient plasticity associated with geometrically necessary dislocations, respectively (Haušild et al., 2025; Liang and Pharr, 2022; Maier et al., 2015). In strain-rate-controlled spherical indentation, however, the two become coupled, because the indentation strain is defined through the evolving contact geometry and the loading protocol therefore determines the indentation depth, contact radius and plastic-zone size together with the loading rate. In the present experiments this coupling was compounded by the different final depths reached at the three nominal strain rates. The measured differences in load, pile-up, residual strain and GND density between these conditions therefore could not, on their own, be attributed to rate sensitivity.

By prescribing loading rate and depth independently, the CPFE simulations separate these contributions. Within the experimental strain-rate range ($3.2 \times 10^{-5}$ to $3.2 \times 10^{-3}$ $s^{-1}$) and within the adopted constitutive framework, strain rate is found to have little effect at a given depth, changing the predicted peak load by less than 1%, whereas depth is seen to control the size of the plastic zone and hence the load, pile-up and extent of the deformation field. Rate becomes significant only at much higher strain rates, where the reduced time for thermal activation raises the flow stress and reduces pile-up. The depth dependence alone reproduces the magnitude of the pile-up changes measured at the different strain rate conditions expriementally, including the larger change in irradiated tungsten (Section 5.4.3). These conclusions are specific to the strain-rate range and constitutive framework examined here. They nevertheless have a general implication: in strain-rate-controlled spherical indentation, a difference of only about 10% in final depth can mimic more than an order of magnitude change in strain rate (Section 5.4.3), so rate-sensitivity parameters extracted from such tests must account for the evolving deformation geometry and for differences in final depth between tests.

## 6.2 How irradiation reorganises plastic deformation

The combined measurements and simulations show that neutron irradiation changes where plastic deformation occurs around the indent. In the irradiated cases, AFM shows pile-up 1.5–2.3 times higher than in unirradiated tungsten, with steeper flanks (Section 5.2). HR-EBSD shows a residual strain field that peaks closer to the indent and falls off far more steeply (Section 5.3). In the simulations, the GND field is more compact after irradiation. Together, these observations indicate that irradiation-induced obstacles restrict dislocation motion and the lateral expansion of the plastic zone, confining deformation to a smaller volume near the indent and forcing more of the displaced material to the free surface. This is consistent with the deformation localisation and dislocation channelling reported for irradiated tungsten and other irradiated metals (Das et al., 2019a; Fournier et al., 2009; Klimenkov et al., 2016; Long et al., 2016).

The larger retained elastic strain in irradiated tungsten may also reflect back-stresses developed against the dense obstacle field. Dislocations held up at loops, voids and Re/Os-rich precipitates would resist reverse motion on unloading, limiting the relaxation of the lattice and leaving steeper residual strain fields around the indent. The present measurements do not isolate this effect, and it is offered as a plausible contribution.

The depth dependence of the irradiation effect on pile-up (Section 5.4.2) can be understood at two levels. Within the homogeneous CPFE representation, it may be arising from the interaction between the expanding plastic zone and the irradiation-dependent constitutive response. At shallow depths, the plastic zone is small, and strain-gradient (GND) hardening, which acts in both materials, contributes a larger fraction of the flow resistance; the irradiation-specific contribution is therefore partially masked by the indentation size effect, and the relative difference between the materials is smaller. As the depth increases, a larger volume of material follows the irradiation-specific hardening and softening response, promoting strain localisation and confinement of deformation near the indent, and hence a steeper increase in pile-up in the irradiated material. In the real microstructure, a corresponding transition may be expected from the discrete nature of the defects. The loops, voids and Re/Os-rich precipitates in the present material may be having spacings of tens to hundreds of nanometres, so a shallow indent samples relatively few obstacles and its response is dominated by the strain gradients imposed by the indenter. As the plastic zone expands, it may be sampling an increasingly representative volume of defects, and the deformation may shift from being gradient-dominated to obstacle-dominated, allowing the athermal resistance and channel-forming tendencies of the irradiated microstructure to develop fully. However, since the model considers the irradiated microstructure as homogeneous, it cannot demonstrate these sampling effects that may be arising from the discrete, spatially correlated nature of the defects. A similar trend, with modest irradiation effects for shallow indents but substantially greater hardening and pile-up once depths exceed several hundred nanometres, has been reported for ion-irradiated Fe and Fe–Cr alloys (Song et al., 2024; Yadav et al., 2026). A practical consequence is that shallow indents may underestimate the effect of irradiation on surface deformation: the predicted ratio of irradiated to unirradiated pile-up rises from about 1.6 at 200 nm to about 2 at depths of 500 nm or more. The irradiation effect on load, in contrast, is approximately independent of depth over the range simulated.

### 6.3 Rate sensitivity and the rate-controlling mechanism after irradiation

The simulations predict a weaker rate dependence for irradiated tungsten at high strain rates: about 9% increase in peak load at $3.2 \times 10^{1}$ s$^{-1}$, compared with about 30–40% for unirradiated tungsten (Section 5.4.1). Physically, this may be understood as a change in the composition of the flow resistance. In unirradiated tungsten, plastic flow is limited by the intrinsic Peierls barrier to screw dislocation glide, which is overcome by thermally activated kink-pair nucleation and is therefore strongly rate sensitive. In irradiated tungsten, the dense population of voids, loops and Re/Os-rich precipitates adds a large, largely athermal resistance, so the rate-sensitive contribution may be forming a smaller fraction of the total. In the present model this outcome is built in, since $\tau_H^0$ enters the critical resolved shear stress as an athermal term. Thus, this result may be interpreted as consistent with the formulation rather than as an independent finding.

More generally, an unchanged rate-controlling mechanism is assumed in the present framework, through the same activation energy and fixed mobile and statistically stored dislocation densities for both material states. The agreement with the measured load–displacement response, pile-up and principal features of the residual strain fields show that this assumption is sufficient to reproduce the experiments, rather than demonstrating that it is uniquely required. Within that caveat, the results are consistent with irradiation redistributing where plastic deformation occurs, rather than changing how it occurs: plasticity remains governed by thermally activated screw dislocation motion, while irradiation adds a largely athermal resistance through dislocation–defect interactions.

### 6.4 Broader implications

Building on previously established constitutive descriptions for irradiation-modified tungsten (Das et al., 2020; S. Das et al., 2018), the present study extends physically based crystal plasticity to strain-rate-dependent deformation of neutron-irradiated material. Calibrated at a single strain rate and assessed against four independent observables, the rate-sensitive framework enables controlled numerical experiments that are not accessible experimentally, allowing the contributions of constitutive rate sensitivity and evolving deformation geometry to be separated. This provides a basis for interpreting strain-rate-dependent nanoindentation of irradiated materials and for developing constitutive descriptions for their transient loading. More broadly, the results show that rate sensitivity cannot be inferred directly from strain-rate-controlled spherical nanoindentation without accounting for the evolving deformation geometry, a conclusion likely to apply to other materials in which strain-gradient plasticity and thermally activated deformation operate together.

## 7 Conclusion

Understanding rate-dependent deformation in neutron-irradiated tungsten is essential for developing constitutive descriptions applicable to the transient loading conditions expected in future fusion reactors. The simultaneous evolution of indentation depth, contact geometry and plastic-zone development, during spherical nanoindentation provides a unique, dynamic probe for investigating the coupling of length-scale and rate effects. In this study, strain-rate-controlled spherical nanoindentation, atomic force microscopy, high-resolution electron backscatter diffraction and a rate-sensitive, strain-gradient crystal plasticity framework were combined to separate the effects of loading rate and deformation length scale in unirradiated and neutron-irradiated single-crystal tungsten. The principal conclusions are as follows.

- In the present experiments, tests performed to the same final indentation strain at nominal strain rates of $3.2 \times 10^{-5}$ to $3.2 \times 10^{-3}$ $s^{-1}$ reached final depths between 336 and 538 nm, so the differences measured between strain-rate conditions could not be attributed to rate sensitivity from the experiments alone.
- A single CPFE framework, with the irradiation-induced resistance estimated from the TEM measured irradiation-defect populations and calibrated at the highest strain rate only, reproduced the experimentally measured response at the two lower strain rates without further parameter adjustment. It captured the load–displacement response, the four-fold pile-up morphology, and the sign and symmetry of the residual lattice strain and GND fields.
- Simulations in which loading rate and indentation depth were varied independently show that, within the experimentally considered strain-rate range and the adopted constitutive framework, strain rate has little effect at a given depth. Between $3.2 \times 10^{-5}$ and $3.2 \times 10^{-3}$ $s^{-1}$, the predicted peak load changes by less than 1% and the peak pile-up by less than 2 nm; a significant rate effect appears only at strain rates well above the experimental range. Indentation depth, in contrast, controls the plastic-zone size and hence the load and pile-up.
- The differences observed between the nominal strain-rate conditions in the experiments therefore arise predominantly from the different final depths reached, and these depth differences are most likely instrumental or control-related rather than a consequence of material rate sensitivity. More generally, a difference of about 10% in final depth produces a change in load equivalent to more than an order of magnitude in strain rate, so final depths must be closely matched, to within about 5–7% for this material, before differences between strain-rate conditions can be attributed to rate sensitivity.
- Neutron irradiation localises plastic deformation around the indent. Pile-up is 1.5–2.3 times higher than in unirradiated tungsten, the residual strain field peaks closer to the indent and decays more steeply, and the simulated GND field is more compact. The irradiation effect on pile-up increases with depth, the ratio of irradiated to unirradiated pile-up rising from about 1.6 at 200 nm to about 2 at 500 nm and beyond, whereas the relative effect on load (about 30–50%) is approximately independent of depth. Shallow indents may therefore underestimate the effect of irradiation on surface deformation.

- The observations are consistent with irradiation changing where plastic deformation occurs rather than the rate-controlling mechanism, which remains the thermally activated glide of screw dislocations.
- More generally, rate sensitivity cannot be inferred directly from strain-rate-controlled spherical nanoindentation without accounting for the evolving deformation geometry, and in particular for differences in final depth between tests. The experimentally assessed, rate-sensitive framework presented here provides a means of separating these contributions in irradiated materials and a basis for constitutive descriptions of irradiated tungsten under transient loading.

**CRediT authorship contribution statement**

**Prashil S. Joshi:** Conceptualization, Methodology, Software, Validation, Formal analysis, Investigation, Writing - original draft. **Bethany Jim:** Conceptualization, Methodology, Investigation, Formal analysis, Writing - review & editing. **Chris Hardie:** Conceptualization, Methodology, Writing – review & editing, Supervision, Funding acquisition. **Angus Wilkinson:** Conceptualization, Writing – review & editing, Supervision, Funding acquisition. **David EJ Armstrong:** Conceptualization, Writing – review & editing, Supervision, Funding acquisition. **Suchandrima Das:** Conceptualization, Methodology, Writing – review & editing, Supervision, Funding acquisition.

**Declaration of competing interests**

The authors declare that they have no known competing financial interests or personal relationships that could have influenced the work reported in this paper.

**Acknowledgements**

The computational work reported in this study was supported by the Indian Institute of Science through the Start-up Research Grant (Grant No. CAPT-24-00003). This work has been part-funded by the EPSRC Fusion Grant 2022/27 [grant number EP/W006839/1]. The authors gratefully acknowledge the Ministry of Education, Government of India, for fellowship support to Prashil S. Joshi. The authors gratefully acknowledge the Centre for Doctoral Training (CDT) for fellowship support to Bethany Jim.

**Data availability**

Data can be made available upon request.

**Appendix A:**

Fig. A.1 compares the measured and predicted pile-up profiles along the ⟨110⟩ direction for both materials at the three nominal strain rates. The measured profiles are averages of line profiles extracted from two diametrically opposite lobes within ±5° of ⟨110⟩, and the same averaged profile was used for calibration at the highest strain rate. The CPFE simulations reproduce the four-lobed pile-up morphology, the alignment of the lobes along ⟨110⟩, and the decrease in pile-up between the nominal strain-rate conditions. For unirradiated tungsten, the predicted pile-up heights agree with the measurements to within about 10 nm at all three strain rates, and the residual depths to within about 35 nm. For irradiated tungsten, the measured lobes are asymmetric at the two faster strain rates, with one lobe up to about 90 nm higher than the opposite one, whereas the simulated profiles are symmetric. The predicted pile-up heights lie within the range spanned by the two measured lobes at these rates, and are about 15 nm higher than measured at the slowest rate. The asymmetry likely reflects local heterogeneity in the irradiated microstructure which is not represented in the model, and it motivated the use of the averaged profile for calibration. The largest difference in residual depth, about 70 nm, occurs for irradiated tungsten at the fastest strain rate.

As the constitutive parameters were calibrated at the highest strain rate only and applied unchanged to the other conditions, these comparisons show that the model reproduces the pile-up response at the experimentally achieved depths, including the larger pile-up of the irradiated material. Because the predicted rate sensitivity in this range is small (Section 5.4.1), the agreement at the lower strain rates primarily reflects the depth-dependent response rather than providing an independent test of the rate formulation (Section 5.4.3).

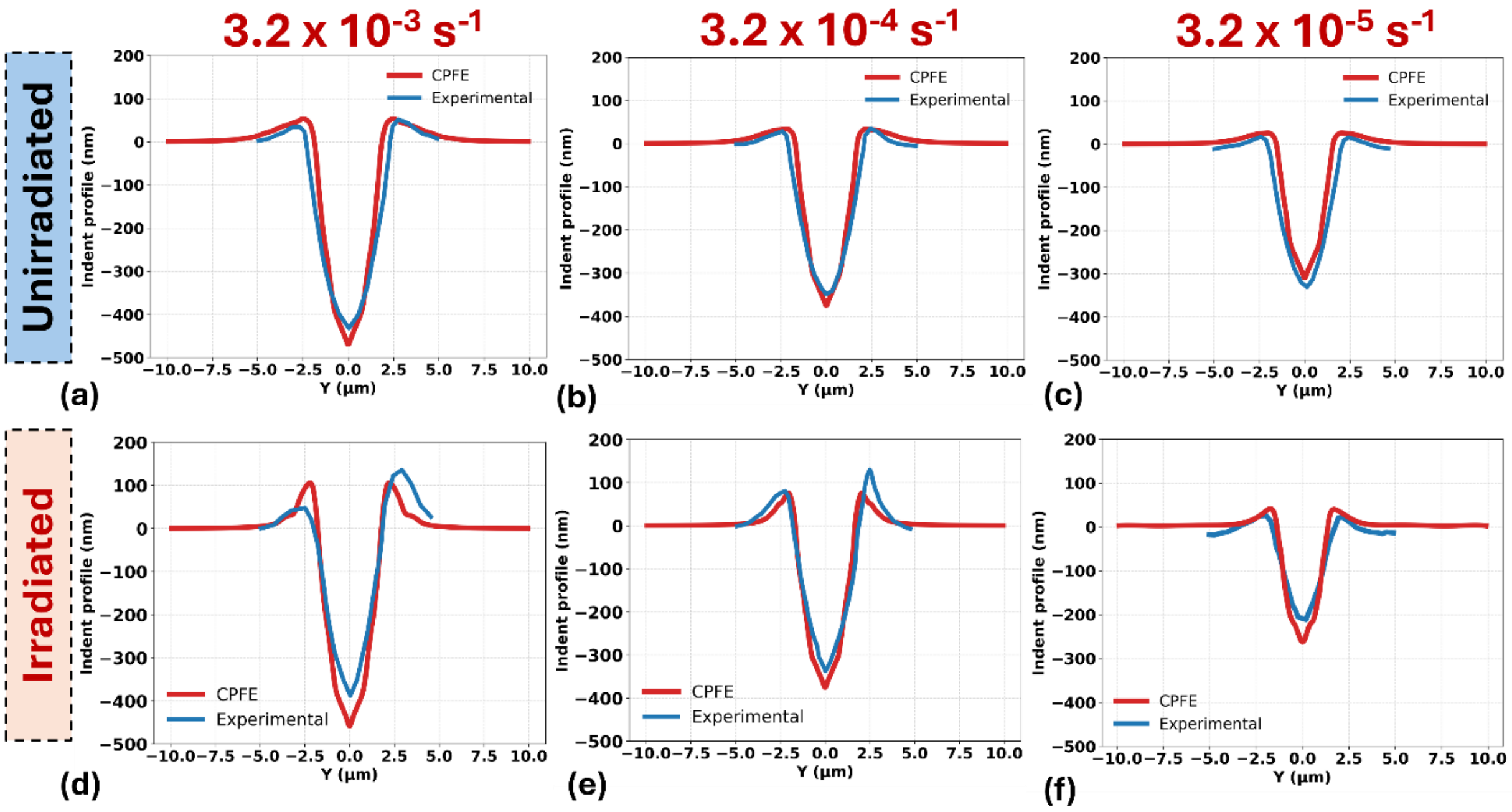

**Fig. A.1.** Line profiles of nano-indents depicting the pile-up in (a) pure tungsten and in (b) neutron-irradiated tungsten as predicted by CPFE and measured by AFM after indentation for strain rates 3.2 x $10^{-3}$ $s^{-1}$, 3.2 x $10^{-4}$ $s^{-1}$, and 3.2 x $10^{-5}$ $s^{-1}$.

## Appendix B:

Fig. B.1 compares the $\varepsilon_{xx}$, $\varepsilon_{xy}$ and $\varepsilon_{yy}$ profiles measured by HR-EBSD with the CPFE predictions, along horizontal lines through the indent centre, for both materials at the three nominal strain rates. HR-EBSD data are not available for irradiated tungsten at the slowest strain rate, so only the CPFE profiles are shown for that condition. Within the indent cavity, the measured profiles show sharp fluctuations where pattern quality is poor, and these are not used for comparison.

Outside the indent, the CPFE reproduces the sign and spatial pattern of all three components: compressive $\varepsilon_{xx}$ and tensile $\varepsilon_{yy}$ on either side of the indent, and an antisymmetric $\varepsilon_{xy}$ profile. Quantitative agreement is better for irradiated tungsten, where the predicted $\varepsilon_{xx}$ captures the position and shape of the compressive trough adjacent to the indent, with a magnitude somewhat larger than measured. For unirradiated tungsten, the model overestimates the residual strain outside the indent considerably, predicting compressive $\varepsilon_{xx}$ of about $-5 \times 10^{-3}$ where about $-1 \times 10^{-3}$ is measured. In both materials, the predicted tensile $\varepsilon_{yy}$ is broader and larger than measured.

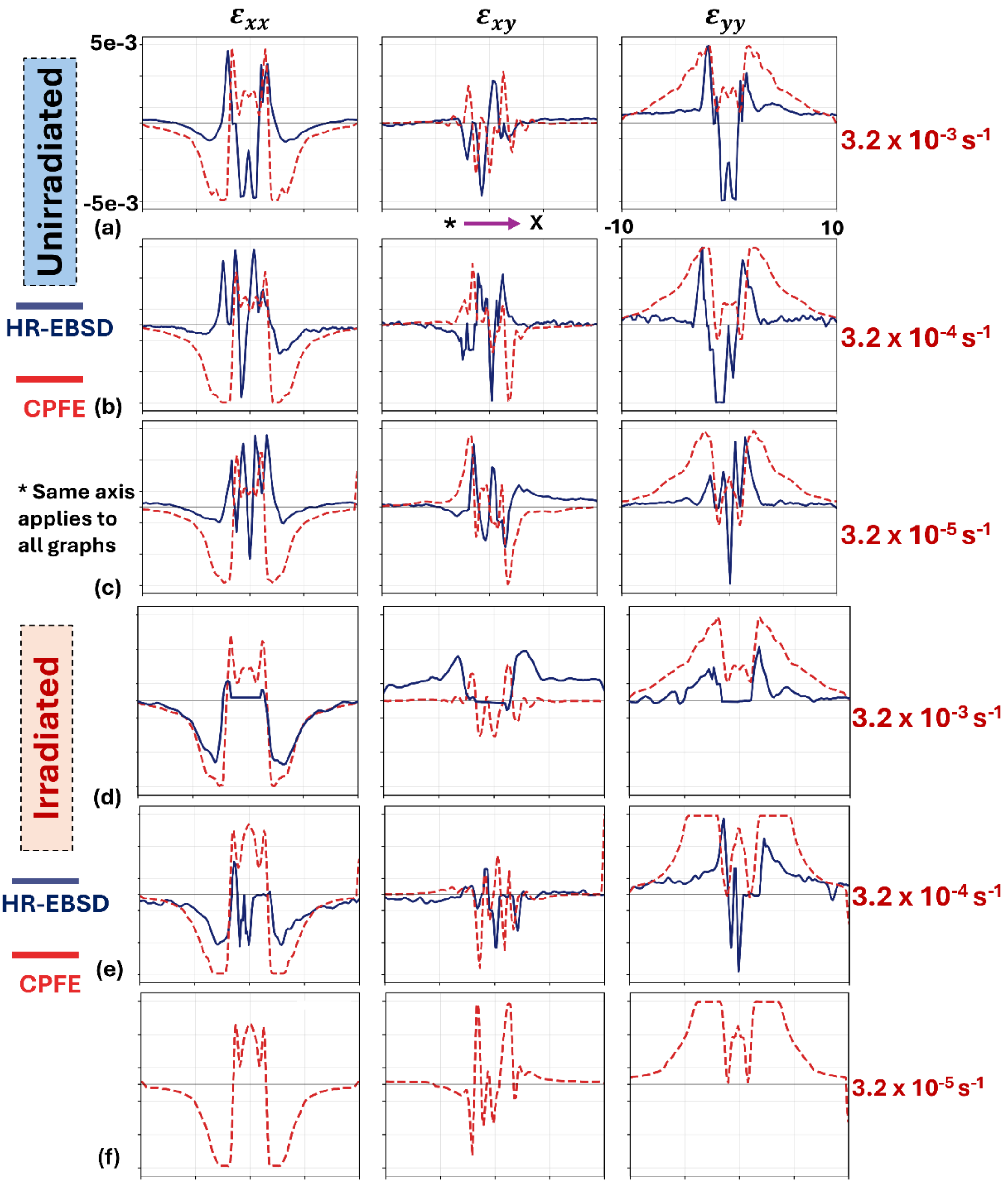

**Fig. B.1**. Comparison of horizontal elastic-strain profiles obtained from HR-EBSD and CPFE for unirradiated and neutron-irradiated tungsten at strain rates of 3.2 x $10^{-3}$ $s^{-1}$, 3.2 x $10^{-4}$ $s^{-1}$, and 3.2 x $10^{-5}$ $s^{-1}$. The columns correspond to the $\varepsilon_{xx}$, $\varepsilon_{xy}$ and $\varepsilon_{yy}$ strain components, respectively. HR-EBSD profiles are shown in navy blue, while CPFE predictions are shown as red dashed lines. The profiles are extracted horizontally through the indent centre, and the vertical axis represents elastic strain, with a common range of -5 x $10^{-3}$ to +5 x $10^{-3}$ for all profiles.

## Appendix C:

| | | |
|---|---|---|
| Elastic modulus $E$ | 421 GPa | |
| Shear modulus $G$ | 164.4 GPa | (Ayres et al., 1975; Bolef and De Klerk, 1963; Featherston and Neighbours, 1963b; Klein and Cardinale, 1993) |
| Poisson's ratio $v$ | 0.28 | (Ayres et al., 1975; Bolef and De Klerk, 1963; Featherston and Neighbours, 1963b; Klein and Cardinale, 1993) |
| Burgers vector $b$ | $2.7 \times 10^{-10}$ m | (Dutta and Dayal, 1963) |
| Activation energy $\Delta F$ | $3.4559 \times 10^{-20}$ J | (Gumbsch et al., 1998) |
| Boltzmann constant $k$ | $1.381 \times 10^{-23}$ J / K | (Sweeney et al., 2013) |
| Temperature $T$ | 298 K | Room temperature assumed during experiments |
| Attempt frequency $\nu$ | $1 \times 10^{11}$ s$^{-1}$ | (Sweeney et al., 2013) |
| Density of statistically stored dislocations $\rho_{SSD}$ | $1 \times 10^{10}$ m$^{-2}$ | *1 |
| Density of mobile dislocations $\rho_{\mathrm{m}}$ | $1 \times 10^{10}$ m$^{-2}$ | *1 |
| Probability of pinning $\Psi$ | 0.1 | Value chosen and kept fixed |
| Critical resolved shear stress $\tau_{c_0}$ | 300 MPa | Fitted to experimental data of unirradiated sample |
| Geometric hardening factor $C'$ | 0.42 | Fitted to experimental data of unirradiated sample |
| Initial value of solute hardening from defects $\tau_H^0$ | 662 MPa | Fitted to experimental data of irradiated sample |
| $\gamma_s$ | 0.12 | Fitted to experimental data of irradiated sample |

**Table. C.1:** Mechanical properties for tungsten from literature and CRSS parameters used for fitting load-displacement, morphological and deformation data obtained from nanoindentation, AFM and HR-EBSD experiments.

***1** *It is evident from the GND density maps shown in Results section, the background density measured by HR-EBSD was on the order of ~$10^{13}$ $m^{-2}$. In the present CPFE analysis, it is assumed that both statistically stored dislocation density ($\rho_{SSD}$) and the mobile dislocation density ($\rho_m$) are lower than the estimated GND density ($\rho_{GND}$) and of the order of ~$10^{10}$ $m^{-2}$.*

## Appendix D: Conversion of indentation strain rate to indenter displacement rate

### D.1 Conversion and choice of representative strain

In the experiments, the indentation strain rate was controlled through feedback (Section 2.3), whereas the CPFE simulations prescribe the indenter displacement rate. For an ideal spherical indenter of radius $R$, and with $h \approx h_c$, the two are related by Eq. (7), which for a fixed strain $\varepsilon$ can be written as

$$\dot{h} = K(\varepsilon)\dot{\varepsilon}, \quad K(\varepsilon) = \frac{576\pi^2 R\varepsilon}{(16 + 9\pi^2\varepsilon^2)^2} \qquad \text{(D.1)}$$

$K$ depends on the strain at which the conversion is evaluated, which increased from 0.02 to 0.1 during the experiments and is spatially heterogeneous beneath the indenter. A single representative strain, $\varepsilon_{rep}$ = 0.008, was therefore adopted (Section 4.1). As Table D.1 shows, the choice of $\varepsilon_{rep}$ changes the prescribed displacement rates by about an order of magnitude between $\varepsilon$ = 0.008 and 0.1. However, because the strain rate enters Eq. (D.1) only as a multiplier, any choice of $\varepsilon_{rep}$ preserves the ten-fold ratios between the three nominal strain rates. If the effective strain of the experiments, $\varepsilon_{eff}$, differs from $\varepsilon_{rep}$, all three displacement rates are in error by the same factor $f = K(\varepsilon_{eff})/K(\varepsilon_{rep})$.

| ε | K(ε) (μm) | $\dot{h}$ at 3.2 × $10^{-3}$ $s^{-1}$ (nm $s^{-1}$) | $\dot{h}$ at 3.2 × $10^{-5}$ $s^{-1}$ (nm $s^{-1}$) |
|---|---|---|---|
| 0.008 (adopted) | 0.888 | 2.84 | 0.0284 |
| 0.02 | 2.21 | 7.08 | 0.0708 |
| 0.05 | 5.40 | 17.3 | 0.173 |
| 0.10 | 9.97 | 31.9 | 0.319 |

**Table D.1.** Conversion factor K(ε) from Eq. (D.1) for R = 5 μm, and the corresponding displacement rates at the fastest and slowest nominal strain rates.

**D.2 Effect of a common rate factor on the flow stress**

In the regime of thermally activated glide, $(|\tau| - \tau_c)\Delta V \gg kT$, so that $\sinh x \approx ½ \exp(x)$, and the slip law (Eq. 9) can be inverted to give the resolved shear stress required to sustain a slip rate $\dot{\gamma}$:

$$\tau \approx \tau_c + \frac{\Delta F}{\Delta V} + \frac{kT}{\Delta V} ln(\frac{2\dot{\gamma}}{\rho_m b^2 \nu}) \tag{D.2}$$

The required stress thus depends on the logarithm of the slip rate. In displacement-controlled indentation, the deformation state at a given depth is governed primarily by the indentation geometry. Multiplying the displacement rate by *f* therefore reaches the same depth, and to first order the same deformation state, in a time shorter by the factor *f*, so that the local slip rates at every material point are multiplied by approximately the same factor. By Eq. (D.2), this raises the flow stress by

$$\Delta\tau = \frac{kT}{\Delta V} ln\, f \tag{D.3}$$

at all strain rates. The GND hardening (Eq. 11) and the irradiation-induced resistance (Eq. 13) depend on the accumulated deformation rather than on time, and the elastic response is rate-independent, so these are unaffected to first order. The difference in flow stress between two strain rates, $(kT/\Delta V) \ln(\dot{\varepsilon}_1/\dot{\varepsilon}_2)$, therefore does not depend on $\varepsilon_{rep}$.

Since $\tau_{c0}$ enters $\tau_c$ additively (Eqs. 11 and 15) and was calibrated at the reference strain rate using the same conversion, the calibration absorbs the offset $\Delta\tau$. The simulations at the two lower strain rates, performed with the same parameters and the same conversion, are therefore not biased by the choice of $\varepsilon_{rep}$.

### D.3 Assumptions

The argument rests on three approximations: the exponential regime of the slip law, in which the offset $(kT/\Delta V)\ln f$ is itself small when $kT/\Delta V$ is small; a uniform rescaling of the local slip rates with the displacement rate, which neglects second-order changes in the distribution of stress and slip beneath the indenter; and the idealised relation of Eq. (7), with $h \approx h_c$. Within these approximations, $\varepsilon_{rep}$ affects the absolute displacement rates prescribed in the simulations but not the comparison between strain-rate conditions on which the conclusions of this work are based.

**References:**

Abernethy, R.G., 2017. Predicting the performance of tungsten in a fusion environment: a literature review. Materials Science and Technology (United Kingdom) 33, 388–399. https://doi.org/10.1080/02670836.2016.1185260

Abernethy, R.G., Gibson, J.S.K.-L., Giannattasio, A., Murphy, J.D., Wouters, O., Bradnam, S., Packer, L.W., Gilbert, M.R., Klimenkov, M., Rieth, M., Schneider, H.-C., Hardie, C.D., Roberts, S.G., Armstrong, D.E.J., 2019. Effects of neutron irradiation on the brittle to ductile transition in single crystal tungsten. Journal of Nuclear Materials 527, 151799. https://doi.org/10.1016/j.jnucmat.2019.151799

Arsenlis, A., Parks, D.M., 1999. Crystallographic aspects of geometrically-necessary and statistically-stored dislocation density. Acta Materialia 47, 1597–1611. https://doi.org/10.1016/S1359-6454(99)00020-8

Ashby, M.F., 1970. The deformation of plastically non-homogeneous materials. Philosophical Magazine 21, 399–424. https://doi.org/10.1080/14786437008238426

Ashraf, F., Castelluccio, G.M., 2023. History and temperature dependent cyclic crystal plasticity model with material-invariant parameters. International Journal of Plasticity 161, 103494. https://doi.org/10.1016/j.ijplas.2022.103494

Ayres, R.A., Shannette, G.W., Stein, D.F., 1975. Elastic constants of tungsten−rhenium alloys from 77 to 298 °K. Journal of Applied Physics 46, 1526–1530. https://doi.org/10.1063/1.321804

Beck, C.E., Hofmann, F., Eliason, J.K., Maznev, A.A., Nelson, K.A., Armstrong, D.E.J., Armstrong, D.E.J., 2017. Correcting for contact area changes in nanoindentation using surface acoustic waves. Scripta Materialia 128, 83–86. https://doi.org/10.1016/j.scriptamat.2016.09.037

Biener, M.M., Biener, J., Hodge, A.M., Hamza, A.V., 2007. Dislocation nucleation in bcc Ta single crystals studied by nanoindentation. Phys. Rev. B 76, 165422. https://doi.org/10.1103/PhysRevB.76.165422

Bolef, D.I., De Klerk, J., 1963. Anomalies in the Elastic Constants and Thermal Expansion of Chromium Single Crystals. Phys. Rev. 129, 1063–1067. https://doi.org/10.1103/PhysRev.129.1063

Britton, T.B., Hickey, J.L.R., 2018. Understanding deformation with high angular resolution electron backscatter diffraction (HR-EBSD). IOP Conference Series: Materials Science and Engineering 304, 12003.

Britton, T.B., Wilkinson, A.J., 2012. High resolution electron backscatter diffraction measurements of elastic strain variations in the presence of larger lattice rotations. Ultramicroscopy 114, 82–95. https://doi.org/10.1016/j.ultramic.2012.01.004
Britton, T.B., Wilkinson, A.J., 2011. Measurement of residual elastic strain and lattice rotations with high resolution electron backscatter diffraction. Ultramicroscopy 111, 1395–1404. https://doi.org/10.1016/j.ultramic.2011.05.007
Bushby, A.J., Roberts, S.G., Hardie, C.D., 2012. Nanoindentation investigation of ion-irradiated Fe-Cr alloys using spherical indenters. Journal of Materials Research 27, 85–90. https://doi.org/10.1557/jmr.2011.304
Byun, T., Hashimoto, N., 2006. Strain Localization in Irradiated Materials. Nuclear Engineering and Technology 38, 619–638.
Choi, I., Dao, M., Suresh, S., 2008. Mechanics of indentation of plastically graded materials—I: Analysis. Journal of the Mechanics and Physics of Solids 56, 157–171. https://doi.org/10.1016/j.jmps.2007.07.007
Coenen, J.W., Arnoux, G., Bazylev, B., Matthews, G.F., Autricque, A., Balboa, I., Clever, M., Dejarnac, R., Coffey, I., Corre, Y., Devaux, S., Frassinetti, L., Gauthier, E., Horacek, J., Jachmich, S., Komm, M., Knaup, M., Krieger, K., Marsen, S., Meigs, A., Mertens, P., Pitts, R.A., Puetterich, T., Rack, M., Stamp, M., Sergienko, G., Tamain, P., Thompson, V., 2015. ELM-induced transient tungsten melting in the JET divertor. Nuclear Fusion 55. https://doi.org/10.1088/0029-5515/55/2/023010
Das, S., Armstrong, D.E.J., Zayachuk, Y., Liu, W., Xu, R., Hofmann, F., 2018. The effect of helium implantation on the deformation behaviour of tungsten: X-ray micro-diffraction and nanoindentation. Scripta Materialia 146, 335–339. https://doi.org/10.1016/J.SCRIPTAMAT.2017.12.014
Das, Suchandrima, Hofmann, F., Tarleton, E., 2018. Consistent determination of geometrically necessary dislocation density from simulations and experiments. International Journal of Plasticity 109, 18–42. https://doi.org/10.1016/j.ijplas.2018.05.001
Das, S., Sand, A., Hofmann, F., 2024. Dislocation pinning in helium-implanted tungsten: A molecular dynamics study. Journal of Nuclear Materials 601, 155293. https://doi.org/10.1016/j.jnucmat.2024.155293
Das, S., Yu, H., Mizohata, K., Tarleton, E., Hofmann, F., 2020. Modified deformation behaviour of self-ion irradiated tungsten: A combined nano-indentation, HR-EBSD and crystal plasticity study. International Journal of Plasticity 135, 102817. https://doi.org/10.1016/j.ijplas.2020.102817
Das, S., Yu, H., Tarleton, E., Hofmann, F., 2019a. Hardening and Strain Localisation in Helium-Ion-Implanted Tungsten. Sci Rep 9, 18354. https://doi.org/10.1038/s41598-019-54753-3
Das, S., Yu, H., Tarleton, E., Hofmann, F., 2019b. Orientation-dependent indentation response of helium-implanted tungsten. Applied Physics Letters 114, 221905. https://doi.org/10.1063/1.5097403
Demir, E., Martinez-Pechero, A., Hardie, C., Tarleton, E., 2024. Restraining geometrically-necessary dislocations to the active slip systems in a crystal plasticity-based finite element framework. International Journal of Plasticity 178, 104013. https://doi.org/10.1016/j.ijplas.2024.104013
Dezerald, L., Proville, L., Ventelon, L., Willaime, F., Rodney, D., 2015. First-principles prediction of kink-pair activation enthalpy on screw dislocations in bcc transition metals: V, Nb, Ta, Mo, W, and Fe. Phys. Rev. B 91, 094105. https://doi.org/10.1103/PhysRevB.91.094105

Dindarlou, S., Castelluccio, G.M., 2022. Substructure-sensitive crystal plasticity with material-invariant parameters. International Journal of Plasticity 155, 103306. https://doi.org/10.1016/j.ijplas.2022.103306
Dümmer, T., Lasalvia, J.C., Ravichandran, G., Meyers, M.A., 1998. Effect of strain rate on plastic flow and failure in polycrystalline tungsten. Acta Materialia 46, 6267–6290. https://doi.org/10.1016/S1359-6454(98)00255-9
Dunne, F.P.E., Rugg, D., Walker, A., 2007. Lengthscale-dependent, elastically anisotropic, physically-based hcp crystal plasticity: Application to cold-dwell fatigue in Ti alloys. International Journal of Plasticity 23, 1061–1083. https://doi.org/10.1016/j.ijplas.2006.10.013
Dürrschnabel, M., Klimenkov, M., Jäntsch, U., Rieth, M., Schneider, H.C., Terentyev, D., 2021. New insights into microstructure of neutron-irradiated tungsten. Scientific reports 11. https://doi.org/10.1038/S41598-021-86746-6
Dutta, B.N., Dayal, B., 1963. Lattice Constants and Thermal Expansion of Palladium and Tungsten up to 878 °C by X-Ray Method. Physica Status Solidi (b) 3, 2253–2259. https://doi.org/10.1002/pssb.19630031207
Featherston, F.H., Neighbours, J.R., 1963a. Elastic Constants of Tantalum, Tungsten, and Molybdenum. Phys. Rev. 130, 1324–1333. https://doi.org/10.1103/PhysRev.130.1324
Featherston, F.H., Neighbours, J.R., 1963b. Elastic Constants of Tantalum, Tungsten, and Molybdenum. Phys. Rev. 130, 1324–1333. https://doi.org/10.1103/PhysRev.130.1324
Federici, G., Andrew, P., Barabaschi, P., Brooks, J., Doerner, R., Geier, A., Herrmann, A., Janeschitz, G., Krieger, K., Kukushkin, A., Loarte, A., Neu, R., Saibene, G., Shimada, M., Strohmayer, G., Sugihara, M., 2003. Key ITER plasma edge and plasma–material interaction issues. Journal of Nuclear Materials 313–316, 11–22. https://doi.org/10.1016/S0022-3115(02)01327-2
Fitzgerald, S.P., 2016. Kink pair production and dislocation motion. Sci Rep 6, 39708. https://doi.org/10.1038/srep39708
Fleck, N.A., Hutchinson, J.W., 1997. Strain Gradient Plasticity, in: Advances in Applied Mechanics. Elsevier, pp. 295–361. https://doi.org/10.1016/S0065-2156(08)70388-0
Fournier, L., Serres, A., Auzoux, Q., Leboulch, D., Was, G.S., 2009. Proton irradiation effect on microstructure, strain localization and iodine-induced stress corrosion cracking in Zircaloy-4. Journal of Nuclear Materials 384, 38–47. https://doi.org/10.1016/j.jnucmat.2008.10.001
Frost, H.J., Ashby, M.F., 1982. Deformation-mechanism maps: the plasticity and creep of metals and ceramics.
Gao, H., Huang, Y., Nix, W.D., Hutchinson, J.W., 1999. Mechanism-based strain gradient plasticity— I. Theory. Journal of the Mechanics and Physics of Solids 47, 1239–1263. https://doi.org/10.1016/S0022-5096(98)00103-3
Gault, B., Chiaramonti, A., Cojocaru-Mirédin, O., Stender, P., Dubosq, R., Freysoldt, C., Makineni, S.K., Li, T., Moody, M., Cairney, J.M., 2021. Atom probe tomography. Nat Rev Methods Primers 1, 51. https://doi.org/10.1038/s43586-021-00047-w
Giannattasio, A., Roberts, S.G., 2007. Strain-rate dependence of the brittle-to-ductile transition temperature in tungsten. Philosophical Magazine 87, 2589–2598. https://doi.org/10.1080/14786430701253197
Giannattasio, A., Yao, Z., Tarleton, E., Roberts, S.G., 2010. Brittle–ductile transitions in polycrystalline tungsten. Philosophical Magazine 90, 3947–3959. https://doi.org/10.1080/14786435.2010.502145
Gilbert, M., 2010. Transmutation and He Production in W and W-alloys. Culham Science Centre.

Gilbert, M.R., Sublet, J.-Ch., 2011. Neutron-induced transmutation effects in W and W-alloys in a fusion environment. Nucl. Fusion 51, 043005. https://doi.org/10.1088/0029-5515/51/4/043005
Gumbsch, P., Riedle, J., Hartmaier, A., Fischmeister, H.F., 1998. Controlling Factors for the Brittle-to-Ductile Transition in Tungsten Single Crystals. Science 282, 1293–1295. https://doi.org/10.1126/science.282.5392.1293
Gussev, M.N., Field, K.G., Busby, J.T., 2015. Deformation localization and dislocation channel dynamics in neutron-irradiated austenitic stainless steels. Journal of Nuclear Materials 460, 139–152. https://doi.org/10.1016/J.JNUCMAT.2015.02.008
Hackett, B.L., Sudharshan Phani, P., Walker, C.C., Oliver, W.C., Pharr, G.M., 2024. On the Strain Rate Sensitivity Measured by Nanoindentation at High Strain Rates. JOM 76, 2936–2945. https://doi.org/10.1007/s11837-023-06338-9
Hardie, C.D., Roberts, S.G., Bushby, A.J., 2015. Understanding the effects of ion irradiation using nanoindentation techniques. Journal of Nuclear Materials 462, 391–401. https://doi.org/10.1016/j.jnucmat.2014.11.066
Hasenhuetl, E., Kasada, R., Zhang, Z., Yabuuchi, K., Kimura, A., 2017. Ion-irradiation effect on strain rate sensitivity of nanoindentation hardness of W single crystal. Materials Transactions 58, 580–586. https://doi.org/10.2320/matertrans.ML201603
Haušild, P., Čech, J., Merle, B., Nohava, J., 2025. Effect of temperature and strain rate on indentation size effect at shallow indentation depths. Materials & Design 255, 114196. https://doi.org/10.1016/j.matdes.2025.114196
Hirth, J.P., Lothe, J., 2016. Theory of dislocations.
Izadpanah Najmabad, S., Olanrewaju, O.F., Pathak, S., Bronkhorst, C.A., Knezevic, M., 2024. Crystal plasticity finite element simulations of nanoindentation and simple compression for yielding of Ta crystals. International Journal of Solids and Structures 300, 112928. https://doi.org/10.1016/j.ijsolstr.2024.112928
Jiang, J., Britton, T.B., Wilkinson, A.J., 2013. Measurement of geometrically necessary dislocation density with high resolution electron backscatter diffraction: Effects of detector binning and step size. Ultramicroscopy 125, 1–9. https://doi.org/10.1016/j.ultramic.2012.11.003
Jim, B., n.d. Micromechanical Testing of Irradiated Tungsten for Fusion. University of Oxford.
Kalidindi, S.R., Pathak, S., 2008. Determination of the effective zero-point and the extraction of spherical nanoindentation stress-strain curves. Acta Materialia 56, 3523–3532. https://doi.org/10.1016/j.actamat.2008.03.036
Klein, C.A., Cardinale, G.F., 1993. Young's modulus and Poisson's ratio of CVD diamond. Diamond and Related Materials 2, 918–923. https://doi.org/10.1016/0925-9635(93)90250-6
Klimenkov, M., Jäntsch, U., Rieth, M., Schneider, H.C., Armstrong, D.E.J., Gibson, J., Roberts, S.G., 2016. Effect of neutron irradiation on the microstructure of tungsten. Nuclear Materials and Energy 9, 480–483. https://doi.org/10.1016/j.nme.2016.09.010
Leitner, A., Maier-Kiener, V., Kiener, D., 2018. Essential refinements of spherical nanoindentation protocols for the reliable determination of mechanical flow curves. Materials & Design 146, 69–80. https://doi.org/10.1016/j.matdes.2018.03.003
Liang, Z.Y., Pharr, G.M., 2022. Decoupling indentation size and strain rate effects during nanoindentation: A case study in tungsten. Journal of the Mechanics and Physics of Solids 165, 104935. https://doi.org/10.1016/j.jmps.2022.104935
Lloyd, M.J., Abernethy, R.G., Gilbert, M.R., Griffiths, I., Bagot, P.A.J., Nguyen-Manh, D., Moody, M.P., Armstrong, D.E.J., 2019. Decoration of voids with rhenium and osmium transmutation products in neutron irradiated single crystal tungsten. Scripta Materialia 173, 96–100. https://doi.org/10.1016/j.scriptamat.2019.07.036
Lloyd, M.J., Haley, J., Jim, B., Abernethy, R., Gilbert, M.R., Martinez, E., Hattar, K., El-Atwani, O., Nguyen-Manh, D., Moody, M.P., Bagot, P.A.J., Armstrong, D.E.J., 2024. Microstructural evolution and transmutation in tungsten under ion and neutron irradiation. Materialia 33, 101991. https://doi.org/10.1016/j.mtla.2023.101991
Lloyd, M.J., London, A.J., Haley, J.C., Gilbert, M.R., Becquart, C.S., Domain, C., Martinez, E., Moody, M.P., Bagot, P.A.J., Nguyen-Manh, D., Armstrong, D.E.J., 2022. Interaction of transmutation products with precipitates, dislocations and grain boundaries in neutron irradiated W. Materialia 22, 101370. https://doi.org/10.1016/j.mtla.2022.101370
Loarte, A., Saibene, G., Sartori, R., Campbell, D., Becoulet, M., Horten, L., Eich, T., Herrmann, A., Matthews, G., Asakura, N., Chankin, A., Leonard, A., Porter, G., Federici, G., Janeschitz, G., Shimada, M., Sugihara, M., 2003. Characteristics of type I ELM energy and particle losses in existing devices and their extrapolation to ITER. Plasma Physics and Controlled Fusion 45, 1549. https://doi.org/10.1088/0741-3335/45/9/302
Long, F., Balogh, L., Brown, D.W., Mosbrucker, P., Skippon, T., Judge, C.D., Daymond, M.R., 2016. Effect of neutron irradiation on deformation mechanisms operating during tensile testing of Zr–2.5Nb. Acta Materialia 102, 352–363. https://doi.org/10.1016/j.actamat.2015.09.032
Luft, A., 1991. Microstructural processes of plastic instabilities in strengthened metals. Progress in Materials Science 35, 97–204. https://doi.org/10.1016/0079-6425(91)90002-B
Maier, V., Durst, K., Mueller, J., Backes, B., Höppel, H.W., Göken, M., 2011. Nanoindentation strain-rate jump tests for determining the local strain-rate sensitivity in nanocrystalline Ni and ultrafine-grained Al. J. Mater. Res. 26, 1421–1430. https://doi.org/10.1557/jmr.2011.156
Maier, V., Schunk, C., Göken, M., Durst, K., 2015. Microstructure-dependent deformation behaviour of bcc-metals – indentation size effect and strain rate sensitivity. Philosophical Magazine 95, 1766–1779. https://doi.org/10.1080/14786435.2014.982741
Makin, M.J., Sharp, J.V., 1965. A Model of Lattice Hardening in Irradiated Copper Crystals with the External Characteristics of Source Hardening. physica status solidi (b) 9, 109–118. https://doi.org/10.1002/pssb.19650090114
Maneiro, M.A.G., Rodríguez, J., 2004. Pile-up effect on nanoindentation tests with spherical–conical tips. https://doi.org/10.1016/j.scriptamat.2004.11.029
Marichal, C., Van Swygenhoven, H., Van Petegem, S., Borca, C., 2013. {110} Slip with {112} slip traces in bcc Tungsten. Sci Rep 3, 2547. https://doi.org/10.1038/srep02547
Miranda, F., Zinovev, A., Poleshchuk, K., Terentyev, D., Yin, C., Verbeken, K., Pardoen, T., 2026. Temperature- and strain rate-dependent plastic deformation of ITER specification tungsten. International Journal of Refractory Metals and Hard Materials 139, 107811. https://doi.org/10.1016/j.ijrmhm.2026.107811
Mishra, S., Das, S., 2026. Evolution of Irradiation-Induced Defect Landscape through Edge Dislocation and Vacancy Loop Interactions in Tungsten. https://doi.org/10.48550/ARXIV.2607.27375
Nečas, D., Klapetek, P., 2012. Gwyddion: an open-source software for SPM data analysis. Open Physics 10, 181–188. https://doi.org/10.2478/s11534-011-0096-2
Nye, J.F., 1953. Some geometrical relations in dislocated crystals. Acta Metallurgica 1, 153–162. https://doi.org/10.1016/0001-6160(53)90054-6

Oliver, W.C., Pharr, G.M., 2004. Measurement of hardness and elastic modulus by instrumented indentation: Advances in understanding and refinements to methodology. Journal of Materials Research 19, 3–20. https://doi.org/10.1557/jmr.2004.19.1.3

Onimus, F., Monnet, I., Béchade, J.L., Prioul, C., Pilvin, P., 2004. A statistical TEM investigation of dislocation channeling mechanism in neutron irradiated zirconium alloys. Journal of Nuclear Materials 328, 165–179. https://doi.org/10.1016/j.jnucmat.2004.04.337

Osetsky, Y., 2022. Strengthening of tungsten by coherent rhenium precipitates formed during low fluence irradiation. Tungsten 4, 20–27. https://doi.org/10.1007/s42864-021-00088-4

Osetsky, Y.N., 2021. Atomic-scale mechanisms of void strengthening in tungsten. Tungsten 3, 65–71. https://doi.org/10.1007/s42864-020-00070-6

Pantleon, W., 2008. Resolving the geometrically necessary dislocation content by conventional electron backscattering diffraction. Scripta Materialia 58, 994–997. https://doi.org/10.1016/j.scriptamat.2008.01.050

Pathak, S., Kalidindi, S.R., Weaver, J.S., Wang, Y., Doerner, R.P., Mara, N.A., 2017. Probing nanoscale damage gradients in ion-irradiated metals using spherical nanoindentation. Scientific reports 7, 11918. https://doi.org/10.1038/s41598-017-12071-6

Pitts, R.A., Carpentier, S., Escourbiac, F., Hirai, T., Komarov, V., Lisgo, S., Kukushkin, A.S., Loarte, A., Merola, M., Naik, A.S., Mitteau, R., Sugihara, M., Bazylev, B., Stangeby, P.C., 2013. A full tungsten divertor for ITER: physics issues and design status. J. Nucl. Mater. 438, S48-56. https://doi.org/10.1016/j.jnucmat.2013.01.008

Rieth, M., Dudarev, S.L., Vicente, S.M.G. de, Aktaa, J., Ahlgren, T., Antusch, S., Armstrong, D.E.J., Balden, M., Baluc, N., Barthe, M.-F., Basuki, W.W., Battabyal, M., Becquart, C.S., Blagoeva, D., Boldyryeva, H., Brinkmann, J., Celino, M., Ciupinski, L., Correia, J.B., Backer, A.D., Domain, C., Gaganidze, E., García-Rosales, C., Gibson, J., Gilbert, M.R., Giusepponi, S., Gludovatz, B., Greuner, H., Heinola, K., Höschen, T., Hoffmann, A., Holstein, N., Koch, F., Krauss, W., Li, H., Lindig, S., Linke, J., Linsmeier, C., López-Ruiz, P., Maier, H., Matejicek, J., Mishra, T.P., Muhammed, M., Muñoz, A., Muzyk, M., Nordlund, K., Nguyen-Manh, D., Opschoor, J., Ordás, N., Palacios, T., Pintsuk, G., Pippan, R., Reiser, J., Riesch, J., Roberts, S.G., Romaner, L., Rosiński, M., Sanchez, M., Schulmeyer, W., Traxler, H., Ureña, A., Laan, J.G. van der, Veleva, L., Wahlberg, S., Walter, M., Weber, T., Weitkamp, T., Wurster, S., Yar, M.A., You, J.H., Zivelonghi, A., 2013. Recent progress in research on tungsten materials for nuclear fusion applications in Europe. Journal of Nuclear Materials 432, 482–500. https://doi.org/10.1016/j.jnucmat.2012.08.018

Song, K., Sheyfer, D., Liu, W., Tischler, J.Z., Das, S., Mizohata, K., Yu, H., Armstrong, D.E.J., Hofmann, F., 2024. Deformation localisation in ion-irradiated Fe and Fe10Cr. Journal of Nuclear Materials 596, 155104. https://doi.org/10.1016/j.jnucmat.2024.155104

Spary, I.J., Bushby, A.J., Jennett, N.M., 2006. On the indentation size effect in spherical indentation, in: Philosophical Magazine. pp. 5581–5593. https://doi.org/10.1080/14786430600854988

Srivastava, K., Gröger, R., Weygand, D., Gumbsch, P., 2013. Dislocation motion in tungsten: Atomistic input to discrete dislocation simulations. International Journal of Plasticity 47, 126–142. https://doi.org/10.1016/j.ijplas.2013.01.014

Sweeney, C.A., Vorster, W., Leen, S.B., Sakurada, E., McHugh, P.E., Dunne, F.P.E., 2013. The role of elastic anisotropy, length scale and crystallographic slip in fatigue crack nucleation. Journal of the Mechanics and Physics of Solids 61, 1224–1240. https://doi.org/10.1016/j.jmps.2013.01.001
Wang, Y., Raabe, D., Klüber, C., Roters, F., 2004. Orientation dependence of nanoindentation pile-up patterns and of nanoindentation microtextures in copper single crystals. Acta Materialia 52, 2229–2238. https://doi.org/10.1016/j.actamat.2004.01.016
Was, G.S., 2016. Fundamentals of radiation materials science: metals and alloys. Springer.
Weinberger, C.R., Boyce, B.L., Battaile, C.C., 2013. Slip planes in bcc transition metals. International Materials Reviews 58, 296–314. https://doi.org/10.1179/1743280412Y.0000000015
Weiss, J., Knezevic, M., 2024. Effects of element type on accuracy of microstructural mesh crystal plasticity finite element simulations and comparisons with elasto-viscoplastic fast Fourier transform predictions. Computational Materials Science 240, 113002. https://doi.org/10.1016/j.commatsci.2024.113002
Weiss, J., Savage, D.J., Knezevic, M., 2024. A parametric study into the influence of Taylor-type scale-bridging artifacts on accuracy of multi-level crystal plasticity finite element models for Mg alloys. Computational Materials Science 232, 112684. https://doi.org/10.1016/j.commatsci.2023.112684
Wilkinson, A.J., 1996. Measurement of elastic strains and small lattice rotations using electron back scatter diffraction. Ultramicroscopy 62, 237–247. https://doi.org/10.1016/0304-3991(95)00152-2
Wilkinson, A.J., Meaden, G., Dingley, D.J., 2006. High-resolution elastic strain measurement from electron backscatter diffraction patterns: New levels of sensitivity. Ultramicroscopy 106, 307–313. https://doi.org/10.1016/j.ultramic.2005.10.001
Wirtz, M., Bardin, S., Huber, A., Kreter, A., Linke, J., Morgan, T.W., Pintsuk, G., Reinhart, M., Sergienko, G., Steudel, I., Temmerman, G.D., Unterberg, B., 2015. Impact of combined hydrogen plasma and transient heat loads on the performance of tungsten as plasma facing material. Nuclear Fusion 55, 123017. https://doi.org/10.1088/0029-5515/55/12/123017
Xu, A., Armstrong, D.E.J., Beck, C., Moody, M.P., Smith, G.D.W., Bagot, P.A.J., Roberts, S.G., 2017. Ion-irradiation induced clustering in W-Re-Ta, W-Re and W-Ta alloys: An atom probe tomography and nanoindentation study. Acta Materialia 124, 71–78. https://doi.org/10.1016/j.actamat.2016.10.050
Yadav, C.B., London, A.J., Tadić, T., Xu, R., Liu, W., Fazinic, S., Das, S., 2026. Characterisation of irradiation damage in Fe–3Cr and Fe–5Cr: A study on the effects of chromium content and temperature. Materials Science and Engineering: A 950, 149464. https://doi.org/10.1016/j.msea.2025.149464
Yang, T.-R., Li, Y.-H., Ren, Q.-Y., Terentyev, D., Xie, H.-X., Gao, N., Zhou, H.-B., Gao, F., Lu, G.-H., 2023. Influence of rhenium-decorated dislocation loops on edge dislocation gliding in tungsten. Scripta Materialia 235, 115624. https://doi.org/10.1016/j.scriptamat.2023.115624
Yi, X., Jenkins, M.L., Briceno, M., Roberts, S.G., Zhou, Z., Kirk, M.A., 2013. *In situ* study of self-ion irradiation damage in W and W–5Re at 500 °C. Philosophical Magazine 93, 1715–1738. https://doi.org/10.1080/14786435.2012.754110
Yuan, Y., Du, J., Wirtz, M., Luo, G.N., Lu, G.H., Liu, W., 2016. Surface damage and structure evolution of recrystallized tungsten exposed to ELM-like transient loads. Nuclear Fusion 56. https://doi.org/10.1088/0029-5515/56/3/036021
Zheng, Z., Balint, D.S., Dunne, F.P.E., 2016. Rate sensitivity in discrete dislocation plasticity in hexagonal close-packed crystals. Acta Materialia 107, 17–26. https://doi.org/10.1016/j.actamat.2016.01.035

Zinkle, S.J., Was, G.S., 2013. Materials challenges in nuclear energy. Acta Materialia 61, 735–758. https://doi.org/https://doi.org/10.1016/j.actamat.2012.11.004